\documentstyle[12pt]{article}
\begin{document}
\begin{flushright}
hep-th/0406127\\ BHU-SNB/Preprint
\end{flushright}
\vskip 1.7cm
\begin{center}
{\bf \Large { HODGE DUALITY OPERATION AND ITS PHYSICAL \\
APPLICATIONS ON SUPERMANIFOLDS}}

\vskip 2.5cm

{\bf R.P.Malik} \\ {\it S. N. Bose National Centre for Basic
Sciences,}\\ {\it Block-JD, Sector-III, Salt Lake, Calcutta-700
098, India}\\

\vskip .2cm

and\\

\vskip .2cm

{\it Centre of Advanced Studies, Physics Department,}\\ {\it
Banaras Hindu University, Varanasi-221 005, India}\\ {\bf E-mail
address: malik@bhu.ac.in}

\vskip 2.5cm

\end{center}

\noindent
{\bf Abstract}:
An appropriate definition of the Hodge duality
$\star$ operation on any arbitrary
dimensional supermanifold has been a long-standing problem.
We define a working rule for
the Hodge duality $\star$ operation on the $(2 + 2)$-dimensional
supermanifold parametrized by a couple of even (bosonic) spacetime
variables $x^\mu (\mu = 0 , 1)$ and a couple of odd (fermionic) variables
$\theta$ and $\bar\theta$ of the Grassmann algebra. The Minkowski spacetime
manifold, hidden in
the supermanifold and parametrized by $x^\mu (\mu = 0, 1)$, is chosen to
be a flat manifold on which a two $(1 + 1)$-dimensional (2D) free Abelian
gauge theory, taken as a prototype field theoretical model, is defined.
We demonstrate the applications of the above definition
(and its further generalization) for the discussion of the (anti-)co-BRST
symmetries that exist for the field theoretical models of
2D- and 4D free Abelian gauge theories considered on the
four $(2 + 2)$- and six $(4 + 2)$-dimensional supermanifolds, respectively.\\


\baselineskip=16pt


\newpage

\noindent
{\bf 1 Introduction}\\

\noindent The geometrical superfield formalism is one of the most
intuitive approaches to gain an insight into some of the physical
and mathematical ideas behind the Becchi-Rouet-Stora-Tyutin (BRST)
formalism which plays a very important role in (i) the covariant
canonical quantization of the gauge theories that are endowed with
the first-class constraints in the language of Dirac's
prescription for the classification of constraints (see, e.g.,
[1,2]), (ii) the proof of unitarity of the ``quantum'' gauge
theories at any arbitrary order of perturbative computations (see,
e.g., [3,4,5]), and (iii) providing a deep connection between the
physics of gauge theories with the mathematical ideas behind the
cohomology (see, e.g., [6-9]) of the differential geometry. In the
usual superfield approach [10-17] to the $p$-form ($ p = 1, 2,
3....)$ Abelian gauge theories, defined on the $D$-dimensional
spacetime manifold, a $(p + 1)$-form super curvature $\tilde F^{(p
+ 1)} = \tilde d \tilde A^{(p)}$ is constructed from the super
exterior derivative $\tilde d = dx^\mu
\partial_\mu + d \theta \partial_\theta + d \bar\theta
\partial_{\bar\theta}$ (with $\tilde d^2 = 0$) and the super
$p$-form connection $\tilde A^{(p)}$ on the $(D + 2)$-dimensional
supermanifold parametrized by $D$-number of even (bosonic)
spacetime coordinates $x^\mu (\mu = 0, 1,2......D-1)$ and a couple
of odd (fermionic) elements $\theta, \bar\theta$ (with $\theta^2 =
\bar\theta^2 = 0, \theta\bar\theta + \bar\theta \theta = 0$) of
the Grassmann algebra which constitute the superspace variable
$Z^M = (x^\mu, \theta, \bar \theta)$. This $(p + 1)$-form super
curvature is subsequently equated, due to the so-called
horizontality condition \footnote{This condition is referred to as
the soul-flatness condition by Nakanishi and Ojima [18] which
amounts to setting equal to zero all the Grassmannian components
of the (anti-)symmetric curvature tensor that constitutes the $(p
+ 1)$-form super curvature on the $(D + 2)$-dimensional
supermanifold.}, to the ordinary $(p + 1)$-form curvature $F^{(p +
1)} = d A^{(p)}$ constructed from the ordinary exterior derivative
$d = dx^\mu \partial_\mu$ (with $d^2 = 0$) and the ordinary
$p$-form connection $A^{(p)}$ on the ordinary $D$-dimensional
Minkowskian spacetime manifold parametrized by the bosonic
spacetime variables $x^\mu$ only. This restriction \footnote{The
horizontality condition has also been applied to 1-form 4D
non-Abelian gauge theory where the six $(4 + 2)$-dimensional
2-form super curvature $\tilde F^{(2)} = \tilde d \tilde A^{(1)} +
\tilde A^{(1)} \wedge \tilde A^{(1)}$ is equated with the 4D
ordinary 2-form curvature $F^{(2)} = d A^{(1)} + A^{(1)} \wedge
A^{(1)}$ leading to the derivation of (anti-)BRST symmetry
transformations for the non-Abelian gauge field and the
corresponding (anti-)ghost fields (see, e.g., [18]).} provides the
geometrical origin and interpretation for (i) the nilpotent
(anti-)BRST symmetry transformations (and the corresponding
nilpotent and conserved charges) as the translation generators
$(\partial/\partial\theta)
\partial/\partial\bar\theta$
along the Grassmannian directions of the
supermanifold, (ii) the nilpotency of the above transformations (and
the corresponding nilpotent generators) as a couple of successive translations
(i.e. $(\partial/\partial\theta)^2 = (\partial/\partial\bar\theta)^2 = 0$)
along the Grassmannian directions of the supermanifold, and (iii) the
anticommutativity of the (anti-)BRST transformations (and the corresponding
conserved and nilpotent charges) as the anticommutativity
$(\partial/\partial\theta) (\partial/\partial\bar\theta)
+ (\partial/\partial\bar\theta) (\partial/\partial\theta) = 0$
of the translation generators along the Grassmannian directions of the
$(D + 2)$-dimensional supermanifold.

It is obvious from the above discussions that, in the
horizontality condition, only one (i.e. $(\tilde d) d$) of the
existing three (super) de Rham cohomological operators ($(\tilde
d) d, (\tilde \delta)\delta, (\tilde \Delta)\Delta$) is exploited
for the geometrical interpretations of some of the key properties
associated with the nilpotent (anti-)BRST transformations and the
corresponding conserved charges. To clarify the above notations,
it is worthwhile to be more specific about the de Rham
cohomological operators of the differential geometry defined on an
ordinary spacetime manifold without a boundary. On such a
manifold, the operators $d = dx^\mu \partial_\mu$, $\delta = \pm *
d *$ and $\Delta = (d + \delta)^2$ form a set of de Rham
cohomological operators where $(\delta)d$ are the nilpotent
(co-)exterior derivatives, $\Delta$ is the Laplacian operator and
$*$ is the Hodge duality operation on the manifold. These
operators obey an algebra: $d^2 = \delta^2 = 0, \Delta = \{d,
\delta \}, [\Delta, d] = [\Delta, \delta] = 0$ showing that
$\Delta$ is the Casimir operator for the whole algebra (see, e.g.,
[6-9] for details). It has been a long-standing problem to exploit
the other nilpotent (i.e. $\tilde \delta^2 = 0, \delta^2 = 0$)
mathematical entities $(\tilde \delta)\delta$ of the (super) de
Rham  cohomological operators in the context of the
dual-horizontality condition ($\tilde \delta \tilde A^{(p)} =
\delta A^{(p)}$) and study its consequences on a  $p$-form gauge
theory in the framework of the geometrical superfield approach to
BRST formalism. Here $\tilde \delta = - \star \tilde d \star$ and
$ \delta = - * d *$ are the super co-exterior derivative and
ordinary co-exterior derivative, respectively. The mathematical
symbols $\star$ and $*$ stand for the Hodge duality operations on
the $(D + 2)$-dimensional supermanifold and $D$-dimensional
ordinary manifold, respectively, and the super Laplacian operator
is defined as $\tilde \Delta = (\tilde d + \tilde \delta)^2$. To
tap the mathematical power of $\tilde \delta = - \star \tilde d
\star$, it is clear that the definition of the Hodge duality
$\star$ operation on the $(D + 2)$-dimensional supermanifold is
quite important.

A consistent and systematic definition of the Hodge duality $*$
operation on an ordinary spacetime manifold of any arbitrary
dimensionality is already quite well-known in the literature (see,
e.g., [6-9] for details). In fact, the existence of the totally
symmetric metric tensor and the totally antisymmetric Levi-Civita
tensor on the spacetime manifold plays a crucial role in such a
consistent and systematic definition of the duality operation
($*$). However, such a consistent, precise and elaborate
definition of the Hodge duality $\star$ operation on a
supermanifold, to the best of our knowledge, is not well-known in
the literature (see, e.g., [18-26] for details). The purpose of
our present paper is to provide a working rule for the definition
of the Hodge duality $\star$ operation on the four $(2 + 2)$- and
six $(4 + 2)$-dimensional supermanifolds on which the 2D- and 4D
free 1-form ($A^{(1)} = dx^\mu A_\mu$) Abelian gauge theories are
defined for the derivation of the nilpotent (anti-)co-BRST
symmetry transformations in the framework of superfield approach
to BRST formalism. We exploit this working rule for the definition
of the $\star$ operation in the context of the dual-horizontality
condition ($\tilde \delta \tilde A^{(1)} = \delta A^{(1)}$) where
the action of the super co-exterior derivative $\tilde \delta = -
\star \tilde d \star$ on the super connection 1-form $\tilde
A^{(1)}$ does require, the action of the Hodge duality $\star$
operations (in $\tilde \delta \tilde A^{(1)} = - \star \tilde d
\star \tilde A^{(1)}$) for the derivations of the (anti-)co-BRST
symmetry transformations. To be more precise, for the case of the
4D Abelian gauge theory, defined on the six $(4 + 2)$-dimensional
supermanifold, the $\star$ operation is defined (i) on the super
1-form $\tilde A^{(1)}$ to produce $(\star\; \tilde A^{(1)})$ as a
super 5-form, and subsequently (ii) on the super 6-form $(\tilde d
\star \tilde A^{(1)})$ to produce a super 0-form $(\star \tilde d
\star \tilde A^{(1)})$ to obtain explicitly $\tilde \delta \tilde
A^{(1)} = - \star \tilde d \star \tilde A^{(1)}$. In exactly
similar fashion, the $\star$ operations could be defined for the
2D free Abelian gauge theory, considered on the four $(2 +
2)$-dimensional supermanifold, for the derivation of the nilpotent
(anti-)co-BRST symmetries. Towards the above goals in mind, we
propose, in a systematic manner, the Hodge duality $\star$
operations on all the possible super forms that could be defined
on the $(2 + 2)$-dimensional supermanifold (cf. Section 2.2)) as
well
 as on the $(4 + 2)$-dimensional supermanifold (cf. Section 3.2). These
definitions are subsequently exploited for the derivation of the
nilpotent (anti-)co-BRST symmetries in the framework of superfield
formalism (cf. Sections 2.3 and 3.3 below). Our present study is
essential on three counts. First and foremost, it has been a
long-standing problem to exploit the potential and power of the
(super) co-exterior derivatives $\tilde \delta = - \star \tilde d
\star$ and $\delta = - * d *$ in the context of the derivations of
some specific nilpotent symmetries for the BRST formulation of the
gauge theories. We find that the above (super) cohomological
operators do play a set of decisive roles in the context of the
derivations of the nilpotent (anti-)co-BRST symmetry
transformations for the 2D- and 4D free Abelian gauge theories.
Second, in our recent works [27-32], we have been able to exploit
$(\tilde \delta)\delta$ in the dual-horizontality condition
($\tilde \delta \tilde A^{(1)} = \delta A^{(1)}$) but the precise
expressions for the $\star$ operations on all the super forms,
defined for some suitable supermanifolds, have not yet been
obtained. Finally, our present study {\it might} turn out to be
useful for the discussion of an interacting gauge theory [33,34]
which has been shown to provide (i) the field theoretical model
for the Hodge theory, and (ii) a model for the interacting
topological field theory where topological $U(1)$ field couples
with the matter (Dirac) fields [33,34].

The contents of our present paper are organized as follows.

In Section 2,
we very briefly recapitulate the bare essentials of the (anti-)BRST- and
(anti-)co-BRST symmetry transformations for the 2D free Abelian gauge theory
in the Lagrangian formulation. We also derive the nilpotent (anti-)BRST
symmetry transformations in the framework of superfield formalism by
exploiting the horizontality condition on the $(2 + 2)$-dimensional
supermanifold and provide the geometrical
interpretation for the nilpotent (anti-)BRST charges $Q_{(a)b}$
(cf. Section 2.1). For
the derivation of the (anti-)co-BRST symmetry transformations in the
superfield formalism, we discuss the dual-horizontality condition and define
the Hodge duality $\star$ operation, in a systematic way, for all the
(super)forms defined
on the four $(2 + 2)$-dimensional supermanifold on which a 2D free
Abelian gauge theory is considered. The double Hodge duality $\star$
operations are also defined for all the (super)forms that are supported
by the $(2 + 2)$-dimensional supermanifold.

Section 3 is devoted to (i) a concise
synopsis of the local, covariant, continuous and nilpotent
(anti-)BRST- and non-local, non-covariant, continuous and nilpotent
(anti-)co-BRST symmetry transformations for the free 4D Abelian theory
in the Lagrangian formulation, (ii) a brief discussion for the derivation
of the (anti-)BRST symmetry transformations in the usual superfield formalism
and its key points of differences with such a derivation for the 2D free
Abelian theory,
(iii) a systematic definition of the single Hodge duality $\star$ operation
(and the double Hodge duality $\star$ operations) for all the (super)forms
defined on the six $(4 + 2)$-dimensional supermanifold, and (iv) the derivation
of the nilpotent (anti-)co-BRST symmetries by exploiting the
$\star$ operation in the context of the dual-horizontality condition.

Finally, in Section 4, we make some concluding remarks.\\

\noindent
{\bf 2 (Anti-)BRST- and (anti-)co-BRST
symmetries for 2D theory: a brief sketch}\\

\noindent
Let us begin with the BRST- and anti-BRST invariant
Lagrangian density ${\cal L}_b$ for the two $(1 + 1)$-dimensional
\footnote{We adopt here the convention and notations such that the flat
2D Minkowskian spacetime manifold is endowed with the flat metric
$\eta_{\mu\nu} =$ diag $(+1, -1)$ and $\Box = \eta^{\mu\nu}\partial_\mu
\partial_\nu = (\partial_0)^2 - (\partial_1)^2,
(\partial \cdot A) = \partial_0 A_0 - \partial_1 A_1, F_{01} = E = -
\varepsilon^{\mu\nu} \partial_\mu A_\nu = - F^{01}$.
Here the 2D antisymmetric Levi-Civita
tensor is chosen to satisfy $\varepsilon_{01} = + 1 = - \varepsilon^{01},
\varepsilon^{\mu\nu} \varepsilon_{\nu\lambda} = \delta^\mu_\lambda$, etc.,
and the Greek indices $\mu, \nu, \lambda...= 0 , 1$ correspond to the
time- and space directions on the 2D flat Minkowskian
spacetime manifold, respectively.}
(2D) free Abelian gauge theory  in the Feynman gauge [3,4,18,35]
$$
\begin{array}{lcl}
{\cal L}^{(2)}_b &=& -
{\displaystyle \frac{1}{4}}\; F^{\mu\nu} F_{\mu\nu} + B (\partial \cdot A)
+ {\displaystyle \frac{1}{2}}\;
 B^2 - i \partial_\mu \bar C \partial^\mu C, \nonumber\\
&\equiv&
{\displaystyle  \frac{1}{2}}\; E^2 + B (\partial \cdot A)
+ {\displaystyle \frac{1}{2}}\; B^2 - i \partial_\mu \bar C \partial^\mu C,
\end{array} \eqno(2.1)
$$ where $F_{\mu\nu} = \partial_\mu A_\nu - \partial_\nu A_\mu$ is
the antisymmetric field strength (curvature) tensor derived from
the 2-form $F^{(2)} = d A^{(1)} = \frac{1}{2} (dx^\mu \wedge
dx^\nu) F_{\mu\nu}$. The latter is constructed by the application
of the exterior derivative $d = dx^\mu \partial_\mu$ (with $ d^2 =
0$) on the 1-form connection $A^{(1)} = dx^\mu A_\mu$ which
defines the vector potential $A_\mu$ for the Abelian gauge theory.
Thus, the operation of $d$ on 1-form increases the degree by $+1$.
It will be noted that $F_{\mu\nu}$ has only electric component and
the magnetic component of $F_{\mu\nu}$ is zero in 2D. The
Nakanishi-Lautrup auxiliary field $B$ has been introduced to
linearize the gauge-fixing term $- \frac{1}{2} (\partial \cdot
A)^2$ and the fermionic $(\bar C^2 = C^2 = 0, C \bar C + \bar C C
= 0$) (anti-)ghost fields $(\bar C)C$ are required to maintain the
unitarity and quantum gauge (i.e. BRST) invariance together for a
given physical process allowed by the theory. At this stage, it is
worth emphasizing that the gauge-fixing term $(\partial\cdot A)$
owes its origin to the other nilpotent $(\delta^2 = 0$)
cohomological operator $\delta$ because the operation of the
latter ($ \delta A^{(1)} = - * d * A^{(1)} = (\partial \cdot A))$
on the 1-form $A^{(1)}$ produces it. The operator $\delta = - * d
*$, which decreases the degree of a form by $1$,  is known as the
co-exterior derivative and $*$ is the Hodge duality operation on
the 2D spacetime manifold. The action of the Laplacian operator
$\Delta$ on the 1-form $A^{(1)}$ (i.e. $\Delta A^{(1)} = dx^\mu
\Box A_\mu$) leads to the derivation of the equation of motion
$\Box A_\mu = 0$ for the gauge field $A_\mu$ if we demand the
validity of the Laplace equation $\Delta A^{(1)} = 0$. The degree
of a form remains intact under the operation of $\Delta$. Thus, we
note that all the three de Rham cohomological operators ($d,
\delta, \Delta$) of differential geometry play very important
roles in the description of the gauge theories. One can linearize
the kinetic energy term $\frac{1}{2} E^2$  of (2.1) by introducing
another auxiliary field ${\cal B}$ as $$
\begin{array}{lcl}
{\cal L}^{(2)}_B &=&  {\cal B} E - \frac{1}{2} {\cal B}^2 + B (\partial \cdot A)
+ \frac{1}{2} B^2 - i \partial_\mu \bar C \partial^\mu C.
\end{array} \eqno(2.2)
$$
For the special case of 2D free Abelian gauge theory, the auxiliary
field ${\cal B}$ is analogous to the Nakanishi-Lautrup field $B$. In fact, the
former linearizes of the kinetic energy term $\frac{1}{2} E^2$ in
exactly the same manner as the latter linearizes
the gauge-fixing term $- \frac{1}{2} (\partial \cdot A)^2$. The above
Lagrangian density (2.2) is endowed with the following local, off-shell
nilpotent $(s_{(a)b}^2 = 0)$ and  anticommuting
($s_b s_{ab} + s_{ab} s_b = 0$) (anti-)BRST $s_{(a)b}$
transformations
\footnote{We follow here the notations adopted in refs. [3,35]. In its
full blaze of glory, the nilpotent $(\delta_B^2 = 0$)
BRST transformations $\delta_B$ are the product of an anticommuting
$(\eta C + C \eta = 0, \eta \bar C + \bar C \eta = 0)$ spacetime
independent parameter $\eta$ and $s_b$ as: $\delta_B = \eta s_b$
where the nilpotency property is carried by
$s_b$ (with $s_b^2 = 0$).}
$$
\begin{array}{lcl}
s_b A_\mu &=& \partial_\mu C, \qquad s_b C = 0, \qquad
s_b \bar C = i B, \quad s_b B = 0, \qquad s_b {\cal B} = 0, \quad
s_b E = 0, \nonumber\\
s_{ab} A_\mu &=& \partial_\mu \bar C, \;\quad s_{ab} \bar C = 0, \quad
s_{ab} C = - i B, \quad s_{ab} B = 0, \quad s_{ab} {\cal B} = 0,
\quad s_{ab} E = 0.
\end{array} \eqno(2.3)
$$ The key point to be noted, at this stage, is the fact that the
kinetic energy term (more precisely the electric field itself),
owing its origin to the exterior derivative $d$ and $A^{(1)}$,
remains invariant under the (anti-)BRST transformations. In
contrast, under the following local, off-shell nilpotent
($s_{(a)d}^2 = 0$) and anticommuting ($s_d s_{ad} + s_{ad} s_d =
0$) (anti-)co-BRST (or (anti-)dual-BRST) transformations
$s_{(a)d}$ $$
\begin{array}{lcl}
s_d A_\mu &=& -\varepsilon_{\mu\nu} \partial^\nu \bar C,
\quad s_d \bar C = 0, \quad
s_d C = - i {\cal B}, \nonumber\\
 s_d B &=& 0, \quad s_d {\cal B} = 0, \quad
s_d (\partial \cdot A) = 0,
\nonumber\\
s_{ad} A_\mu &=& -\varepsilon_{\mu\nu} \partial^\nu C,
\quad s_{ad}  C = 0, \quad
s_{ad} \bar C = + i {\cal B}, \nonumber\\
s_{ad} B &=& 0, \qquad s_{ad} {\cal B} = 0,
\qquad s_{ad} (\partial \cdot A) = 0,
\end{array} \eqno(2.4)
$$ it is the gauge-fixing term (more precisely $(\partial\cdot A)$
itself), owing its origin to the co-exterior derivative $\delta$
and $A^{(1)}$, remains invariant. The anticommutator $(s_w = \{
s_b, s_d \} = \{s_{ab}, s_{ad} \})$ of the (anti-)BRST- and
(anti-)co-BRST transformations leads to the existence of a
non-nilpotent $s_w^2 \neq 0$ bosonic symmetry transformation in
the theory [36-38] under which the (anti-)ghost fields do not
transform at all. This bosonic symmetry is the analogue of the
Laplacian operator of the differential geometry. There exists a
global ghost scale symmetry transformation: $ s_g A_\mu = 0, s_g B
= 0, s_g {\cal B} = 0, s_g C = -\Lambda C, s_g \bar C = +\Lambda
\bar C$, under which, the Lagrangian density (2.2) remains
invariant. Here $\Lambda$ is an infinitesimal spacetime
independent (global) parameter. All the above six symmetry
transformations can  be concisely expressed, in terms of the
generic local field $\Sigma (x) =  A_\mu (x), C (x), \bar C (x), B
(x), {\cal B} (x)$, as $$
\begin{array}{lcl}
s_r\; \Sigma (x) = - i \; [\; \Sigma (x) , Q_r\; ]_{\pm}, \qquad\;\;
r = b, ab, d, ad, w, g,
\end{array} \eqno(2.5)
$$
where $(+)-$ signs on the square brackets stand for the (anti-)commutator
for the generic local field $\Sigma$ being (fermionic)bosonic in nature.
Here $Q_r$ are the generator of transformations which can be derived from
the Noether's theorem. Their exact form is not required for our present
discussion but their explicit and exact form can be found in [36-38].\\

\noindent
{\bf 2.1 Superfield formulation of (anti-)BRST symmetries: a concise review}\\

\noindent We begin here with a four ($2 + 2$)-dimensional
supermanifold parametrized by the superspace coordinates $Z^M =
(x^\mu, \theta, \bar \theta)$ where $x^\mu (\mu = 0, 1)$ are the
two even (bosonic) spacetime coordinates and $\theta$, $\bar
\theta$ are the two odd (Grassmannian) coordinates (with $\theta^2
= \bar \theta^2 = 0, \theta \bar \theta + \bar \theta \theta =
0)$. On this supermanifold, one can define a supervector 1-form
superfield $\tilde A^{(1)} = d Z^M \tilde A_M (x, \theta,
\bar\theta)$ with the following component multiplet superfields
(see, e.g., [13,12]) \noindent $$
\begin{array}{lcl}
\tilde A_M (x,\theta,\bar\theta)  = \Bigl ( B_{\mu} (x, \theta, \bar \theta),
\;\Phi (x, \theta, \bar \theta),
\;\bar \Phi (x, \theta, \bar \theta) \Bigr ).
\end{array} \eqno(2.6)
$$
It will be noted that component superfields
$B_\mu (x,\theta,\bar\theta), \Phi (x,\theta,\bar\theta),
\bar \Phi (x, \theta, \bar\theta)$ are the generalization of the basic
local fields $A_\mu (x), C (x), \bar C (x)$, defined on the 2D ordinary
spacetime manifold, to the four $(2 + 2)$-dimensional supermanifold.
The most general expansion of these superfields
along the Grassmannian directions of the supermanifold, is [13,27-34]
$$
\begin{array}{lcl}
B_{\mu}\; (x, \theta, \bar \theta) &=& A_{\mu} (x)
+ \;\theta\;  \bar R_\mu (x)
+  \;\bar \theta\;  R_\mu (x)
+ i \;\theta \;\bar \theta \;S_{\mu} (x), \nonumber\\
\Phi\; (x, \theta, \bar \theta) &=& C (x) \;+ i \; \theta\;  \bar B (x)
\;- i \;\bar \theta\;  {\cal B} (x)
\;+ i \;\theta \;\bar \theta \; s (x),
 \nonumber\\
\bar \Phi\; (x, \theta, \bar \theta) &=& \bar C (x)
\;- i \;\theta\;\bar {\cal B} (x)) \;+ i\;\bar \theta\;  B (x)
\;+ i\;\theta\;\bar \theta \; \bar s (x),
\end{array} \eqno(2.7)
$$
where $(+)-$ signs in the above expansion have been chosen for the algebraic
convenience. It should be noted that (i) in the limit $\theta \to 0,
\bar\theta \to 0$, we get back the local basic fields $A_\mu (x), C (x),
\bar C (x)$ of the theory from the superfields
$B_\mu (x,\theta,\bar\theta), \Phi (x,\theta,\bar\theta),
\bar \Phi (x,\theta, \bar\theta)$. (ii) The fermionic degrees of freedom
($ C, \bar C, R_\mu, \bar R_\mu, s, \bar s)$ match with that of the
bosonic $(A_\mu, S_\mu, B, \bar B, {\cal B}, \bar {\cal B})$ degrees of
freedom so that the expansion can be consistent with the basic tenets
of supersymmetry. (iii) All the fields on the r.h.s. of the expansion
are the local functions of spacetime $x^\mu$ alone.

The secondary fields (i.e. $R_\mu, \bar R_\mu, S_\mu, s, \bar s$)
can be expressed in terms of the basic fields (i.e. $A_\mu, C,
\bar C, B, {\cal B}$) of the Lagrangian density (2.2) by
exploiting the horizontality condition ($\tilde F^{(2)} =
F^{(2)}$) where the super curvature 2-form $\tilde F^{(2)} =
\tilde d \tilde A^{(1)}$, defined on the $(2 + 2)$-dimensional
supermanifold, is equated with the ordinary 2-form curvature
$F^{(2)} = d A^{(1)}$, defined on the 2D ordinary flat Minkowskian
spacetime manifold. The explicit expressions for these forms are
$$
\begin{array}{lcl}
\tilde F^{(2)} =  \frac{1}{2}\; (d Z^M \wedge d Z^N)\; \tilde
F_{MN} = \tilde d \tilde A^{(1)}, \qquad
 F^{(2)} =  \frac{1}{2}\; (dx^\mu \wedge d x^\nu)\;
F_{\mu\nu} = d A^{(1)},
\end{array} \eqno(2.8)
$$ where the exact expressions for $\tilde d$, $\tilde A^{(1)}$
and $\tilde d \tilde A^{(1)} = \tilde F^{(2)}$ (constructed by
$\tilde d, \tilde A^{(1)}$),  are $$
\begin{array}{lcl}
\tilde d &=& \;d Z^M \;\partial_{M} = d x^\mu\; \partial_\mu\; +
\;d \theta \;\partial_{\theta}\; + \;d \bar \theta
\;\partial_{\bar \theta}, \nonumber\\ \tilde A^{(1)} &=& d Z^M\;
\tilde A_{M} = d x^\mu \;B_{\mu} (x , \theta, \bar \theta) + d
\theta\; \bar \Phi (x, \theta, \bar \theta) + d \bar \theta\; \Phi
( x, \theta, \bar \theta),
\end{array}\eqno(2.9)
$$
$$
\begin{array}{lcl}
\tilde F^{(2)} &=& \tilde d \tilde A^{(1)} = (d x^\mu \wedge d
x^\nu)\; (\partial_{\mu} B_\nu) - (d \theta \wedge d \theta)\;
(\partial_{\theta} \bar \Phi) + (d x^\mu \wedge d \bar \theta)
(\partial_{\mu} \Phi - \partial_{\bar \theta} B_{\mu})\nonumber\\
&+& (d x^\mu \wedge d \theta) (\partial_{\mu} \bar \Phi -
\partial_{\theta} B_{\mu}) - (d \bar\theta \wedge d \bar\theta)
(\partial_{\bar\theta} \Phi) - (d \theta \wedge d \bar\theta)
(\partial_{\theta} \Phi + \partial_{\bar\theta} \bar \Phi).
\end{array}\eqno(2.10)
$$ Ultimately, the application of soul-flatness (horizontality)
condition ($ \tilde d \tilde A^{(1)} = d A^{(1)}$), leads to the
following restrictions (cf. (2.11) ) and thereby the ensuing
relationships (cf. (2.12)) $$
\begin{array}{lcl}
\partial_\mu \Phi &=& \partial_{\bar\theta} B_\mu, \qquad
\partial_\mu \bar \Phi = \partial_{\theta} B_\mu, \qquad
\partial_{\theta} \bar \Phi = \partial_{\bar\theta} \Phi = 0, \nonumber\\
\partial_\mu R_\nu &=& \partial_{\nu } R_\mu, \qquad
\partial_\mu \bar R_\nu = \partial_{\nu } \bar R_\mu, \qquad
\partial_{\theta} \Phi  + \partial_{\bar\theta} \bar \Phi = 0,
\end{array} \eqno(2.11)
$$
$$
\begin{array}{lcl}
R_{\mu} \;(x) &=& \partial_{\mu}\; C(x), \qquad
\bar R_{\mu}\; (x) = \partial_{\mu}\;
\bar C (x), \qquad
S_{\mu}\; (x) = \partial_{\mu} B\; (x),
\nonumber\\
{\cal B}\; (x) &=&
\bar {\cal B}\; (x) = 0,
 \qquad \bar s\;(x) = s\; (x) = 0, \qquad B (x) + \bar B (x) = 0.
\end{array} \eqno(2.12)
$$
The insertion of all the above values into the most general expansion (2.7)
on the $(2 + 2)$-dimensional supermanifold
leads to the derivation of the off-shell nilpotent
(anti-)BRST transformations for the most basic fields $A_\mu, C, \bar C$
as expressed below
$$
\begin{array}{lcl}
B_{\mu}\; (x, \theta, \bar \theta) &=& A_{\mu} (x)
+ \;\theta\; (s_{ab} A_{\mu} (x))
+ \;\bar \theta\; (s_{b} A_{\mu} (x))
+ \;\theta \;\bar \theta \;(s_{b} s_{ab} A_{\mu} (x)), \nonumber\\
\Phi\; (x, \theta, \bar \theta) &=& C (x) \;+ \; \theta\; (s_{ab} C (x))
\;+ \;\bar \theta\; (s_{b} C (x))
\;+ \;\theta \;\bar \theta \;(s_{b}\; s_{ab} C (x)),
 \nonumber\\
\bar \Phi\; (x, \theta, \bar \theta) &=& \bar C (x)
\;+ \;\theta\;(s_{ab} \bar C (x)) \;+\bar \theta\; (s_{b} \bar C (x))
\;+\;\theta\;\bar \theta \;(s_{b} \;s_{ab} \bar C (x)).
\end{array} \eqno(2.13)
$$
It should be noted that (i) the third- and the fourth terms in the
above expansion of $\Phi (x,\theta,\bar\theta)$ and the second- and the fourth
terms of the above expansion of $\bar\Phi (x,\theta,\bar\theta)$ are
exactly equal to zero because $s_b C = 0, s_{ab} \bar C = 0, s_{(a)b} B = 0$.
(ii) A comparison with (2.5) establishes the geometrical interpretation
for the nilpotent ($Q_{(a)b}^2 = 0$) (anti-)BRST charges $Q_{(a)b}$
as the translation generators along $(\theta)\bar\theta$-directions
of the supermanifold. In fact, there exists a mapping
$$
\begin{array}{lcl}
s_{b} \leftrightarrow \mbox{Lim}_{\theta \to 0}
{\displaystyle \frac{\partial}{\partial \bar\theta}} \leftrightarrow Q_b,
\qquad
s_{ab} \leftrightarrow \mbox{Lim}_{\bar \theta \to 0}
{\displaystyle \frac{\partial}{\partial \theta}} \leftrightarrow Q_{ab},
\end{array} \eqno(2.14)
$$
among the (anti-)BRST transformations $s_{(a)b}$, the translation
generators along $(\theta)\bar\theta$-direction of the supermanifold
and the nilpotent (anti-)BRST charges $Q_{(a)b}$. (iii) The geometrical
interpretation of the nilpotency property is contained in the
translations generators which satisfy $(\partial/\partial\theta)^2 =
(\partial/\partial\bar\theta)^2 = 0$. (iv) The anticommutativity
properties of the transformations $s_b s_{ab} + s_{ab} s_b = 0$
and their corresponding generators $Q_b Q_{ab} + Q_{ab} Q_b = 0$ are
reflected in the specific property of the translation generators
$(\partial/\partial\theta) (\partial/\partial\bar\theta)
+ (\partial/\partial\bar\theta) (\partial/\partial\theta) = 0$.
(v) Under the (anti-)BRST transformations, the superfields
$(\bar\Phi)\Phi$ convert themselves from  the general superfields
(cf. (2.7)) to the (anti-)chiral superfields (i.e.
$\Phi (x,\theta,\bar\theta) = C (x) - i\; \theta\; B (x),
\bar \Phi (x,\theta,\bar\theta) = \bar C (x) + i\; \bar \theta \;B (x)$)
because these satisfy $\partial_{\bar\theta} \Phi = 0,
\partial_\theta \bar \Phi = 0$.\\

\noindent
{\bf 2.2 Hodge duality operation on $(2 + 2)$-dimensional supermanifold}\\

\noindent It is evident from the previous Section that we have
been able to derive the local, covariant, nilpotent ($s_{(a)b}^2 =
0$) and anticommuting ($s_b s_{ab} + s_{ab} s_b = 0$) (anti-)BRST
symmetry transformations $s_{(a)b}$ without any recourse to the
definition of the Hodge duality operation. This is primarily due
to the fact that we have exploited only the (super) exterior
derivatives $(\tilde d) d$ and the (super) 1-form connections
$(\tilde A^{(1)}) A^{(1)}$ in the horizontality condition $\tilde
d \tilde A^{(1)} = d A^{(1)}$ where the Hodge duality operation
plays {\it no} role at all. For the derivation of the local,
covariant, nilpotent ($s_{(a)d}^2 = 0$) and anticommuting ($s_d
s_{ad} + s_{ad} s_d = 0$) (anti-)co-BRST symmetry transformations
$s_{(a)d}$, we have to tap the potential and power of the super
co-exterior derivative $\tilde \delta = - \star \tilde d \star$
and the ordinary co-exterior derivative $\delta = - * d *$ in the
dual-horizontality condition $\tilde \delta \tilde A^{(1)} =
\delta A^{(1)}$, where (i) $\star$ and $*$ are the Hodge duality
operations, and (ii) $\tilde A^{(1)}$ and $A^{(1)}$ are the
(super) 1-form connections on the supermanifold and ordinary
manifold, respectively. On the four $(2 + 2)$-dimensional
supermanifold, there exist three independent 4-forms (and their
linear combinations are also allowed). These independent 4-forms
are $$
\begin{array}{lcl}
&&\phi_{1} = \frac{1}{2!}
(dx^\mu \wedge dx^\nu \wedge d \theta \wedge d \bar\theta)\;
{\cal F}_{\mu\nu\theta\bar\theta}, \quad
\phi_{2}  = \frac{1}{2!}
(dx^\mu \wedge dx^\nu \wedge d \theta \wedge d \theta)\;
{\cal F}_{\mu\nu\theta\theta}, \nonumber\\
&&\phi_{3} = \frac{1}{2!}
(dx^\mu \wedge dx^\nu \wedge d \bar\theta \wedge d \bar\theta)\;
{\cal F}_{\mu\nu\bar\theta\bar\theta}.
\end{array} \eqno(2.15)
$$
It will be noted that (i) the wedge product between
the pure Grassmannian differentials is symmetric (i.e.
$d\theta \wedge d\theta = d \theta \wedge d\theta,
d\theta \wedge d\bar \theta = d \bar \theta \wedge d\theta,
d\bar \theta \wedge d\bar \theta = d \bar \theta \wedge d\bar \theta$),
the wedge product between the pure spacetime differentials
is antisymmetric (i.e. $dx^\mu \wedge dx^\nu = - dx^\nu \wedge dx^\mu$),
and the wedge product between the mixed differentials is also antisymmetric
(i.e $dx^\mu \wedge d\theta = - d \theta \wedge dx^\mu,
dx^\mu \wedge d\bar\theta = - d \bar\theta \wedge d x^\mu$). Accordingly, the
covariant indices of ${\cal F}$'s will also be symmetric as well as
antisymmetric corresponding to our specific choice of these indices.
(ii) On the $(2 + 2)$-dimensional supermanifold, more than two spacetime-
as well as two Grassmannian differentials (e.g. $dx^\mu \wedge dx^\nu
\wedge dx^\lambda \wedge d \theta, dx^\mu \wedge d \theta
\wedge d \theta \wedge d \bar\theta$ etc.)
are not allowed. (iii) For the present supermanifold, the overall numerical
factors (e.g. $\frac{1}{2!}$), present in the definition of the superforms
(e.g. (2.15)), correspond to such numerical factors present in the definition
of ordinary forms on the ordinary spacetime manifold. (iv) The Hodge duality
$\star$ operation for some selected super-forms on a six $(4 + 2)$-dimensional
supermanifold have been defined in our
earlier work [42]. However, some ad-hoc assumptions have been made in [42].
No such assumptions have been made in our present
Hodge duality $\star$ definitions.
(v) The operation
of the Hodge duality on a given form does not affect ${\cal F}$'s {\it per se}.
However, the wedge products, present in the above forms,
are affected by the Hodge
duality operation. For instance,
a single  Hodge duality $\star$ operation on the wedge
product of the above cited differentials of the 4-forms,
on the $(2 + 2)$-dimensional supermanifold, is
$$
\begin{array}{lcl}
&&\star\; (dx^\mu \wedge dx^\nu \wedge d \theta \wedge d \bar\theta)
= \varepsilon^{\mu\nu}, \qquad
\star\; (dx^\mu \wedge dx^\nu \wedge d \theta \wedge d \theta)
= \varepsilon^{\mu\nu} \; s^{\theta\theta}, \nonumber\\
&&\star\; (dx^\mu \wedge dx^\nu \wedge d \bar \theta \wedge d \bar\theta)
= \varepsilon^{\mu\nu}\; s^{\bar\theta\bar\theta},
\end{array} \eqno(2.16)
$$
which, ultimately, imply the following zero-forms:
$$
\begin{array}{lcl}
\star\; \phi_{1} = \frac{1}{2!}\; \varepsilon^{\mu\nu}\;
{\cal F}_{\mu\nu\theta\bar\theta}, \qquad
\star \; \phi_{2} = \frac{1}{2!}\; \varepsilon^{\mu\nu} s^{\theta\theta}\;
{\cal F}_{\mu\nu\theta\theta}, \qquad
\star \; \phi_{3} = \frac{1}{2!}\;
 \varepsilon^{\mu\nu} s^{\bar\theta\bar\theta}\;
{\cal F}_{\mu\nu\bar\theta\bar\theta}.
\end{array} \eqno(2.17)
$$
At this juncture, a few comments are in order. First, in contrast to the
ordinary spacetime differentials where $(dx^\mu \wedge dx^\mu) = 0$, the
Grassmann differentials of the form $(d\theta \wedge d \theta)$ and
$(d \bar\theta \wedge d \bar\theta)$ are non-zero on the
supermanifold. Second, the coordinates $x^0, x^1, \theta, \bar\theta$
correspond to the {\it four} linearly independent directions on the
($2 + 2$)-dimensional supermanifold. This is why, a single $\star$
operation on ($dx^\mu \wedge dx^\nu \wedge d \theta \wedge d \bar\theta$)
 yields only $\varepsilon^{\mu\nu}$ on the r.h.s. The same does not happen
when we take a single $\star$ operation on
$(dx^\mu \wedge dx^\nu \wedge d \theta \wedge d \theta)$ and
$(dx^\mu \wedge dx^\nu \wedge d \bar\theta \wedge d \bar \theta)$
because $(d\theta \wedge d \theta)$ and $(d \bar\theta \wedge d \bar\theta)$
do not incorporate the linearly independent differentials
 $d\theta$ and $d \bar\theta$ together.
Third, the symmetric quantities
$s^{\theta\theta}$ and $s^{\bar\theta\bar\theta}$ have been introduced
so that one can keep track of the Grassmannian wedge products when
a second Hodge duality operation is applied on a given form. For instance,
two successive $\star$ operations on the wedge products corresponding to
the independent 4-forms, yield the following
$$
\begin{array}{lcl}
&&\star\;
[\;\star\; (dx^\mu \wedge dx^\nu \wedge d \theta \wedge d \bar\theta)\;]
= - \; (dx^\mu \wedge dx^\nu \wedge d \theta \wedge d \bar\theta), \nonumber\\
&&\star\;
[\;\star\; (dx^\mu \wedge dx^\nu \wedge d \theta \wedge d \theta)\;]
= -\;  (dx^\mu \wedge dx^\nu \wedge d \theta \wedge d \theta),
\nonumber\\
&&\star\;
[\;\star\; (dx^\mu \wedge dx^\nu \wedge d \bar \theta \wedge d \bar\theta)\;]
= -\; (dx^\mu \wedge dx^\nu \wedge d \bar \theta \wedge d \bar\theta),
\end{array} \eqno(2.18)
$$
where we have used the following inputs while taking
the second $\star$ operation
$$
\begin{array}{lcl}
&&\star\; \bigl [\;\varepsilon^{\mu\nu}\;\bigr ] =
\frac{1}{2!}\; \varepsilon_{\sigma\rho}\; (dx^\sigma \wedge d x^\rho
\wedge d \theta \wedge d \bar\theta)\;\varepsilon^{\mu\nu}, \nonumber\\
&&\star\;\bigl  [\;\varepsilon^{\mu\nu}\; s^{\theta\theta}\;\bigr ] =
\frac{1}{2!}\; \varepsilon_{\sigma\rho}\; (dx^\sigma \wedge d x^\rho
\wedge d \theta \wedge d \theta)
\;\varepsilon^{\mu\nu}, \nonumber\\
&&\star\;\bigl [\;\varepsilon^{\mu\nu}\; s^{\bar\theta\bar\theta}\;\bigr ]
= \frac{1}{2!}\; \varepsilon_{\sigma\rho}\; (dx^\sigma \wedge d x^\rho
\wedge d \bar \theta \wedge d \bar\theta)\;
\varepsilon^{\mu\nu}.
\end{array} \eqno(2.19)
$$
Thus, it is clear that the presence of the constant symmetric factors
$s^{\theta\theta}, s^{\bar\theta\bar\theta}$ in (2.16) do provide a
kind of guidance for
the operation of a couple of Hodge duality $\star$ operations on a
given wedge product (see, e.g., (2.18) and (2.19)). The
double $\star$ operations are essential because our $\star$ definition
should comply with the general requirements of a duality
invariant theory where $\star (\star G) = \pm G$ is true for any arbitrary
form $G$ (see, e.g., [39]).

Let us concentrate now on the 3-forms. These independent forms are
five in number on the $(2 + 2)$-dimensional supermanifold. These are
as given below
$$
\begin{array}{lcl}
&& \tau_{1} = \frac{1}{2!}\; (dx^\mu \wedge dx^\nu \wedge d \theta) \;
T_{\mu\nu\theta}, \;\;\;\qquad\;\;\;
\tau_{2} = \frac{1}{2!}\; (dx^\mu \wedge dx^\nu \wedge d\bar\theta)\;
T_{\mu\nu\bar\theta}, \nonumber\\
&&
\tau_{3} =
(d x^\mu \wedge d \theta \wedge d\theta)
\; T_{\mu\theta\theta}, \;\;\;\;\;\;\qquad\;\;\;\;\;\;
\tau_{4} =
(d x^\mu \wedge d \bar \theta \wedge d\bar\theta)\;
T_{\mu\bar\theta\bar\theta}, \nonumber\\
&&
\tau_{5} =
(d x^\mu \wedge d  \theta \wedge d\bar\theta)\;
\;T_{\mu\theta\bar\theta}.
\end{array} \eqno(2.20)
$$
As discussed earlier, the operation of the Hodge duality would affect
the wedge products. This is why, we shall obtain a set of 1-forms as the dual
to the above 3-forms. The explicit expressions for a single $\star$ operation
on the wedge products corresponding to 3-forms, are
$$
\begin{array}{lcl}
&& \star\;(dx^\mu \wedge dx^\nu \wedge d \theta) =
\varepsilon^{\mu\nu}\;(d \bar\theta), \;\;\;\;\;\;\qquad\;\;
\star\; (dx^\mu \wedge dx^\nu \wedge d\bar\theta)\; =
\varepsilon^{\mu\nu}\;(d\theta), \nonumber\\ && \star\;(d x^\mu
\wedge d \theta \wedge d\bar\theta) = \varepsilon^{\mu\nu}\;(d
x_\nu), \;\;\;\qquad\;\;\;\; \star\; (d x^\mu \wedge d \bar \theta
\wedge d\bar\theta)\; = \varepsilon^{\mu\nu}\;(d x_\nu)
s^{\bar\theta\bar\theta}, \nonumber\\ && \star\; (d x^\mu \wedge d
\theta \wedge d\theta)\; = \varepsilon^{\mu\nu}\;(d x_\nu)
s^{\theta\theta}.
\end{array} \eqno(2.21)
$$
Application of (2.21) to (2.20) (with inputs as the analogue
of (2.19)) imply
$$
\begin{array}{lcl}
&& \star \;\tau_{1} = \frac{1}{2!}\; \varepsilon^{\mu\nu}\; (d \bar \theta) \;
T_{\mu\nu\theta}, \;\;\;\qquad\;\;\;\;\;\;
\star\;\tau_{2} = \frac{1}{2!}\;
\varepsilon^{\mu\nu}\;(d \theta)\;
T_{\mu\nu\bar\theta}, \nonumber\\
&&
\star\; \tau_{3} = \varepsilon^{\mu\nu} \;s^{\theta\theta}\;
(d x_\nu)\; T_{\mu\theta\theta}, \;\;\;\qquad\;\;\;
\star\; \tau_{4} =
\varepsilon^{\mu\nu}\;s^{\bar\theta\bar\theta}\;(d x_\nu)\;
T_{\mu\bar\theta\bar\theta}, \nonumber\\
&&
\star\; \tau_{5} =
\varepsilon^{\mu\nu}\; (d x_\nu)\;
\;T_{\mu\theta\bar\theta},
\end{array} \eqno(2.22)
$$
which are dual to the 3-forms given in (2.20). The double $\star$
operation on the wedge products corresponding to 3-forms, are
$$
\begin{array}{lcl}
&&\star\;
 [\; \star\; (dx^\mu \wedge dx^\nu \wedge d \theta)\;] = -
(dx^\mu \wedge dx^\nu \wedge d \theta), \nonumber\\
&& \star\;[ \;\star\; (dx^\mu \wedge dx^\nu \wedge d \bar\theta)\;] = -
 (dx^\mu \wedge dx^\nu \wedge d \bar\theta),
\nonumber\\
&& \star\; [\; \star\; (dx^\mu \wedge d \theta \wedge d \bar\theta)\;] =
+\; (dx^\mu \wedge d \theta \wedge d \bar\theta), \nonumber\\
&&\star\; [\; \star\; (dx^\mu \wedge d \theta \wedge d \theta)\; ] =
+ (dx^\mu \wedge d \theta \wedge d \theta), \nonumber\\
&& \star\; [\; \star\; (dx^\mu \wedge d \bar \theta \wedge d \bar \theta)\; ]
= +\;
 (dx^\mu \wedge d \bar \theta \wedge d \bar \theta).
\end{array} \eqno(2.23)
$$
There exist six independent 2-forms on the four $(2 + 2)$-dimensional
supermanifold as
$$
\begin{array}{lcl}
&& \chi_{1}  = \frac{1}{2!}\; (dx^\mu \wedge dx^\nu) \;
S_{\mu\nu}, \;\;\qquad\;
\chi_{2} = (d \theta \wedge d \bar\theta)\;
S_{\theta\bar\theta}, \nonumber\\
&&
\chi_{3} =  (d\theta \wedge d \theta) \;
S_{\theta\theta}, \;\;\;\;\;\qquad\;\;\;\;
\chi_{4} =
(d\bar\theta \wedge d \bar\theta)\;
S_{\bar\theta\bar\theta}, \nonumber\\
&&\chi_{5} =  (dx^\mu \wedge d\theta)
\;S_{\mu\theta}, \;\;\;\qquad \;\;\;\;
\chi_{6} =  (dx^\mu \wedge d\bar\theta)
\;S_{\mu\bar\theta}.
\end{array} \eqno(2.24)
$$
A single $\star$ operation on the wedge products corresponding to the
above 2-forms are as
$$
\begin{array}{lcl}
&& \star\; (dx^\mu \wedge dx^\nu) = \varepsilon^{\mu\nu}\;
(d \theta \wedge d \bar\theta), \;\;\qquad
\star\; (d \theta \wedge d\bar\theta)
= \frac{1}{2!} \varepsilon^{\mu\nu} (dx_\mu \wedge dx_\nu), \nonumber\\
&& \star\; (d \theta \wedge d\theta) = \frac{1}{2!} s^{\theta\theta}
\varepsilon^{\mu\nu} (dx_\mu \wedge dx_\nu), \quad
 \star\; (d \bar\theta \wedge d\bar\theta)
= \frac{1}{2!} s^{\bar\theta\bar\theta}
\varepsilon^{\mu\nu} (dx_\mu \wedge dx_\nu), \nonumber\\
&& \star\;(dx^\mu \wedge d \theta) = \varepsilon^{\mu\nu}\; (dx_\nu \wedge
d \bar\theta), \;\;\qquad\;
\star\;(dx^\mu \wedge d \bar \theta) = \varepsilon^{\mu\nu}\; (dx_\nu \wedge
d \theta),
\end{array} \eqno(2.25)
$$
which clearly establish the fact that the dual of 2-forms (cf. 2.24)
are 2-forms on a four $(2 + 2)$-dimensional supermanifold as listed below
$$
\begin{array}{lcl}
&& \star\; \chi_{1} = \frac{1}{2!}\; \varepsilon^{\mu\nu} \;
(d \theta \wedge d \bar\theta)\;
S_{\mu\nu},\;\; \;\;\;\qquad\;\;\;\;\;\;
\star\; \chi_{2} =  \frac{1}{2!}\; \varepsilon_{\sigma\rho}\;
 (dx^\sigma \wedge dx^\rho)\;
S_{\theta\bar\theta}, \nonumber\\
&&
\star\;\chi_{3} =  s^{\theta\theta} \;
\frac{1}{2!}\;
\varepsilon_{\mu\nu}\; (dx^\mu \wedge dx^\nu)
\;S_{\theta\theta}, \qquad
\star\;\chi_{4} =
s^{\bar\theta\bar\theta}\;
\frac{1}{2!}\;
\varepsilon_{\mu\nu}\; (dx^\mu \wedge dx^\nu)
\;S_{\bar\theta\bar\theta}, \nonumber\\
&&\star\; \chi_{5} = \varepsilon^{\mu\nu}\; (dx_\nu \wedge d \bar\theta)\;
\;S_{\mu\theta}, \;\;\;\qquad \;\;\;\;\;\;\;
\star\;\chi_{6} =  \varepsilon^{\mu\nu} \;(dx_\nu \wedge d\theta)
\;S_{\mu\bar\theta}.
\end{array} \eqno(2.26)
$$
The double $\star$ operation on the wedge products corresponding to
the six independent 2-forms on the $(2 + 2)$-dimensional supermanifold is
$$
\begin{array}{lcl}
&& \star\; [\;\star\; (dx^\mu \wedge dx^\nu)\;] = -\;
(dx^\mu \wedge dx^\nu),\qquad
\star\; [\; \star\; (d \theta \wedge d\bar\theta)\;]
= -\;
 (d \theta \wedge d\bar\theta),
\nonumber\\
&& \star\; [\; \star\; (d \theta \wedge d\theta)] = -\;
 (d \theta \wedge d\theta), \;\;\;\qquad\;\;\;\;\;
\star\; [\; \star\; (d \bar\theta \wedge d\bar\theta)\;] = -\;
(d \bar\theta \wedge d\bar\theta),
\nonumber\\
&& \star\; [\;\star\;(dx^\mu \wedge d \theta)\;] =
 +\; (dx^\mu \wedge d \theta), \;\;\qquad\;\;
\star\; [\;\star\;(dx^\mu \wedge d \bar \theta)\;]  =
+\; (dx^\mu \wedge d \bar \theta).
\end{array} \eqno(2.27)
$$
It is straightforward to guess that there exist only three independent
1-forms on the four $(2 + 2)$-dimensional supermanifold as
\footnote{In general, a set of three 1-forms can be constructed from
the spacetime differential $(dx_\mu)$. These are
$\alpha_1^{(1)} = dx^\mu {\cal A}^{(1)}_\mu,
\alpha_1^{(2)} = dx^\mu s^{\theta\theta} {\cal A}^{(2)}_\mu,
\alpha_1^{(3)} = dx^\mu s^{\bar\theta\bar\theta} {\cal A}^{(3)}_\mu$.
A single $\star$ operation yields
$\star\; (dx^\mu) = \varepsilon^{\mu\nu}
(dx_\nu \wedge d\theta \wedge d \bar\theta),
\star\; [ dx^\mu s^{\theta\theta}]
= \varepsilon^{\mu\nu} (dx_\nu \wedge d\theta \wedge d\theta),
\star\; [ dx^\mu s^{\bar\theta\bar\theta}]
= \varepsilon^{\mu\nu} (dx_\nu \wedge d\bar\theta \wedge d\bar\theta)$. For
obvious reasons, such kind of a triplet of 1-forms can {\it not} be
constructed from the 1-forms $\alpha_2$ as well as  $\alpha_3$ because
their Hodge dual forms are not well defined on a $(2 + 2)$-dimensional
supermanifold.}
$$
\begin{array}{lcl}
\alpha_{1} =
(dx^\mu)\;
{\cal A}_{\mu}, \qquad
\alpha_{2} =
(d \theta)\;
{\cal A}_{\theta}, \qquad
\alpha_{3} =
(d \bar\theta)\;
{\cal A}_{\bar\theta}.
\end{array} \eqno(2.28)
$$
A single $\star$ operation on the above independent 1-forms would
lead to the 3-forms on the four $(2 + 2)$-dimensional supermanifold.
The operation of the single Hodge duality on the
independent 1-form differentials are
$$
\begin{array}{lcl}
&& \star\; (dx^\mu) =  \varepsilon^{\mu\nu}\; (d x_\nu \wedge d \theta
\wedge d \bar\theta), \qquad
\star\; (d \theta) = \frac{1}{2!}\; \varepsilon^{\mu\nu}\;
(dx_\mu \wedge d x_\nu \wedge d \bar\theta), \nonumber\\
&& \star\; (d \bar\theta) = \frac{1}{2!}\; \varepsilon^{\mu\nu} \;
(dx_\mu \wedge d x_\nu \wedge d \theta),
\end{array} \eqno(2.29)
$$
which finally imply the following independent
3-forms corresponding to the independent 1-forms of equation (2.28),
defined on the $(2 + 2)$-dimensional supermanifold, namely;
$$
\begin{array}{lcl}
&&\star\; \alpha_{1} =
\varepsilon^{\mu\nu} \;(dx_\nu \wedge d \theta \wedge d \bar\theta)\;
{\cal A}_{\mu}, \;\;\;\qquad\;\;\;
\star\;\alpha_{2} =
\frac{1}{2!}\; \varepsilon_{\sigma\rho}\;
(dx^\sigma \wedge dx^\rho \wedge d \bar\theta)\;
{\cal A}_{\theta}, \nonumber\\
&&\star\;\alpha_{3} =
\frac{1}{2!}\; \varepsilon_{\sigma\rho}\;
(dx^\sigma \wedge dx^\rho \wedge d \theta)\;
{\cal A}_{\bar\theta}.
\end{array} \eqno(2.30)
$$
The result  of a couple of successive $\star$ operations on the
differentials, corresponding to the 1-forms on the $(2 + 2)$-dimensional
supermanifold, is given by
$$
\begin{array}{lcl}
 \star \; [\;\star\; (dx^\mu)\;] =  +\; (d x^\mu), \qquad
\star \; [\;\star\; (d \theta)\;] =  -\; (d \theta), \qquad
\star \; [\;\star\; (d \bar\theta)\;] =  -\; (d \bar\theta).
\end{array} \eqno(2.31)
$$
We shall be exploiting the above Hodge duality operations on the
wedge products of the differentials of a given form in the forthcoming
Section 2.3
in the context of the derivation of the (anti-)co-BRST symmetries for
the  2D free 1-form Abelian gauge theory considered on
a four $(2 + 2)$-dimensional supermanifold.\\

\noindent
{\bf 2.3 Superfield formulation of (anti-)co-BRST symmetries for 2D theory}\\

\noindent It is clear from the symmetry transformations (2.4) that
the local, covariant, continuous, nilpotent ($s_{(a)d}^2 = 0$) and
anticommuting ($s_d s_{ad} + s_{ad} s_d = 0$) (anti-)co-BRST
symmetries $s_{(a)d}$ exist for the Lagrangian density (2.2)
describing the free (non-interacting) Abelian gauge theory on the
flat 2D Minkowskian spacetime manifold. Exploiting the
dual-horizontality condition $\tilde \delta \tilde A^{(1)} =
\delta A^{(1)}$ with the following inputs $$
\begin{array}{lcl}
\tilde \delta \tilde A^{(1)} = - \star\; \tilde d\; \star \tilde
A^{(1)}, \qquad \delta A^{(1)} = -\; *\; d *\; A^{(1)} = (\partial
\cdot A),
\end{array} \eqno(2.32)
$$ we expect to obtain all the secondary fields of the super
expansion (2.7) in terms of the basic fields of the Lagrangian
density (2.2) of the  theory. Towards this goal in mind, we first
explicitly compute $\tilde \delta \tilde A^{(1)} = - \star \;
\tilde d\; \star \; \tilde A^{(1)}$ taking the help of the
definitions (2.9) and the Hodge duality operations discussed
earlier. First, the dual ($\star \tilde A^{(1)})$ of the super
1-form connection $\tilde A^{(1)} = d Z^M \tilde A_M$ is a 3-form
on the $(2 + 2)$-dimensional supermanifold. The explicit
expression for this 3-form (i.e. dual to the 1-form super
connection $\tilde A^{(1)}$) is $$
\begin{array}{lcl}
\star\; \tilde A^{(1)} &=& \varepsilon^{\mu\nu} \; (dx_\nu \wedge
d \theta \wedge d \bar\theta)\; B_\mu + \frac{1}{2!}\;
\varepsilon_{\sigma\rho}\; (dx^\sigma \wedge dx^\rho \wedge d
\bar\theta)\; \bar \Phi \nonumber\\ &+& \frac{1}{2!}\;
\varepsilon_{\sigma\rho}\; (dx^\sigma \wedge dx^\rho \wedge d
\theta)\;  \Phi,
\end{array} \eqno(2.33)
$$ where we have used the definition of the 1-form super
connection $\tilde A^{(1)}$ from (2.9) and the Hodge duality
operations on the 1-forms from (2.29). We apply now the super
exterior derivative $\tilde d = d Z^M \partial_M$ from (2.9) on
the 3-form dual  super connection (2.33), the outcome is $$
\begin{array}{lcl}
\tilde d\; (\star \tilde A^{(1)}) &=& \varepsilon^{\mu\nu}\;
(dx_\xi \wedge dx_\nu \wedge d\theta \wedge d \bar\theta)\;
(\partial^\xi B_\mu) - \frac{1}{2!} \varepsilon_{\sigma\rho}\;
(dx^\sigma \wedge dx^\rho \wedge d\theta \wedge d\bar\theta)\;
(\partial_\theta \bar \Phi)\nonumber\\ &-&
 \frac{1}{2!} \varepsilon_{\sigma\rho}\;
(dx^\sigma \wedge dx^\rho \wedge d \bar\theta \wedge d\bar\theta)\;
(\partial_{\bar\theta} \bar \Phi)
- \frac{1}{2!} \varepsilon_{\sigma\rho}\;
(dx^\sigma \wedge dx^\rho \wedge d\theta \wedge d \bar \theta)\;
(\partial_{\bar\theta} \Phi)\nonumber\\
&-& \frac{1}{2!} \varepsilon_{\sigma\rho}\;
(dx^\sigma \wedge dx^\rho \wedge d\theta \wedge d \theta)\;
(\partial_\theta \Phi).
\end{array} \eqno(2.34)
$$
A few remarks are in order. First, all the wedge products with more than two
spacetime differentials- as well as Grassmannian differentials are dropped out
because a $(2 + 2$)-dimensional supermanifold cannot support
such forms. Second, the negative signs, in the above, have cropped up because
$(d \theta \partial_\theta) (dx^\mu \wedge dx^\nu \wedge d \theta) \bar \Phi
= - (dx^\mu \wedge dx^\nu \wedge d\theta \wedge d \theta)\;
\partial_\theta \bar\Phi$, etc. The stage is now set for the application
of the $(- \star)$ on the above super 4-forms which will lead to the derivation
of a 0-form (superscalar) on the supermanifold.
Exploiting the Hodge duality operation, defined
in (2.16), we obtain the following expression
$$
\begin{array}{lcl}
\tilde \delta \tilde A^{(1)} = - \star \; \tilde d \; \star \;
\tilde A^{(1)} = (\partial \cdot B) - (\partial_\theta \bar\Phi +
\partial_{\bar \theta} \Phi) - s^{\theta\theta}\; (\partial_\theta
\Phi) - s^{\bar\theta\bar\theta}\; (\partial_{\bar\theta} \bar
\Phi),
\end{array} \eqno(2.35)
$$ where we have used $\varepsilon_{\sigma\rho}
\varepsilon^{\sigma\rho} = - 2!, \varepsilon^{\mu\nu}
\varepsilon_{\xi\nu} = - \delta^\mu_\xi$, etc. When the above
superscalar is equated with the ordinary scalar (i.e. $\delta
A^{(1)} = - * d * A^{(1)} = (\partial \cdot A)$) due to the
requirement of the dual-horizontality condition (i.e. $\tilde
\delta \tilde A^{(1)} = \delta A^{(1)}$), we obtain the following
restrictions $$
\begin{array}{lcl}
(\partial \cdot B) - (\partial_\theta \bar \Phi +
\partial_{\bar\theta} \Phi) = (\partial \cdot A), \qquad
\partial_\theta \Phi = 0, \qquad \partial_{\bar\theta} \bar \Phi = 0.
\end{array} \eqno(2.36)
$$
The insertion of the most general super expansions (cf. (2.7))
on the $(2 + 2)$-dimensional supermanifold for the superfields
$B_\mu (x,\theta,\bar\theta), \Phi (x,\theta,\bar\theta),
\bar \Phi (x,\theta, \bar\theta)$ into the above restriction leads to
$$
\begin{array}{lcl}
&&(\partial \cdot R) (x) = 0, \;\;\qquad\; (\partial \cdot \bar R) (x) = 0,
\;\;\qquad\;
(\partial \cdot S) (x) = 0, \nonumber\\
&& s (x) = \bar s (x) = B (x) = \bar B (x) = 0, \;\;\;\qquad \;\;
{\cal B} (x) + \bar {\cal B} (x) = 0.
\end{array} \eqno(2.37)
$$
It is worthwhile to mention that, unlike the horizontality condition
where the secondary fields are expressed explicitly
and exactly in terms of
the basic fields of the Lagrangian density (2.2), the dual-horizontality
condition provides only the restrictions that are quoted in (2.37). For
the 2D free Abelian gauge theory, the local, covariant and continuous
solutions for the above restrictions exist as given below
\footnote{It will be noted that the non-local and non-covariant solutions to
 the restrictions (2.37) also exist. For the 2D Abelian case,
we have $R_0 = i \bar C, R_1 = i (\partial_0 \partial_1/\nabla^2)
\bar C, \bar R_0 = i C, \bar R_1 = i
(\partial_0\partial_1/\nabla^2) C$, etc. However, for our present
discussions, we avoid such kind of pathological choices. In fact,
for the 4D Abelian theory, this kind of symmetries exist, too
(see, e.g. [40,41], for details).} $$
\begin{array}{lcl}
R_\mu = - \varepsilon_{\mu\nu} \partial^\nu \bar C, \qquad
\bar R_\mu = - \varepsilon_{\mu\nu} \partial^\nu C, \qquad
S_\mu = + \varepsilon_{\mu\nu} \partial^\nu {\cal B}.
\end{array} \eqno(2.38)
$$
Substitution of the above values into the most general super expansion
in (2.7) leads to the following expression
for the expansion {\it vis-{\`a}-vis} the off-shell nilpotent (anti-)co-BRST
transformations of equation (2.4):
$$
\begin{array}{lcl}
B_{\mu}\; (x, \theta, \bar \theta) &=& A_{\mu} (x)
+ \;\theta\; (s_{ad} A_{\mu} (x))
+ \;\bar \theta\; (s_{d} A_{\mu} (x))
+ \;\theta \;\bar \theta \;(s_{d} s_{ad} A_{\mu} (x)), \nonumber\\
\Phi\; (x, \theta, \bar \theta) &=& C (x) \;+ \; \theta\; (s_{ad} C (x))
\;+ \;\bar \theta\; (s_{d} C (x))
\;+ \;\theta \;\bar \theta \;(s_{d}\; s_{ad} C (x)),
 \nonumber\\
\bar \Phi\; (x, \theta, \bar \theta) &=& \bar C (x)
\;+ \;\theta\;(s_{ad} \bar C (x)) \;+\bar \theta\; (s_{d} \bar C (x))
\;+\;\theta\;\bar \theta \;(s_{d} \;s_{ad} \bar C (x)).
\end{array} \eqno(2.39)
$$
This equation
is the analogue of the expansion in (2.13) where (anti-)BRST symmetry
transformations have been derived. It is clear from (2.39) (which
produces the (anti-)co-BRST transformations for the basic fields
$A_\mu, C, \bar C$) that (anti-)co-BRST nilpotent charges $Q_{(a)d}$,
similar to the (anti-)BRST charges $Q_{(a)b}$,
correspond to the translation generators $(\partial/\partial\theta,
\partial/\partial\bar\theta)$ along the Grassmannian directions
$(\theta)\bar\theta$ of the supermanifold. However, there is a clear-cut
distinction between these two sets of charges when it comes to the
discussion of the nilpotent transformations for the (anti-)ghost fields
corresponding to  the fermionic superfields $\Phi$
and $\bar\Phi$. For instance, under the nilpotent anti-BRST transformations,
the superfield $\Phi$ becomes anti-chiral (i.e.
$\Phi = C + \theta\; (s_{ab} C (x))$) but the same superfield
becomes chiral $(i.e. \
Phi = C (x) + \bar\theta \; (s_d C (x))$)
due to the co-BRST transformations. Similar arguments and
interpretations can be provided for the
nature of the superfield $\bar \Phi$ as far as the off-shell nilpotent
BRST and anti-co-BRST transformations are concerned.\\

\noindent
{\bf 3 (Anti-)BRST and (anti-)co-BRST
symmetries for 4D theory: a brief synopsis}\\

\noindent
Let us start off with the analogue of the Lagrangian density (2.1)
for the 4D free Abelian gauge theory defined on the four
dimensional
\footnote{ We follow here the notations and conventions such that
4D Minkowskian manifold is endowed with a flat metric $\eta_{\mu\nu} =$
diag $(+1 , -1, -1, -1)$ and the totally antisymmetric 4D Levi-Civita
tensor $\varepsilon_{\mu\nu\lambda\zeta}$ is chosen to satisfy
$\varepsilon_{0123} = + 1 = - \varepsilon^{0123},
\varepsilon_{0ijk} = \epsilon_{ijk} = - \varepsilon^{0ijk},
\varepsilon_{\mu\nu\lambda\zeta} \varepsilon^{\mu\nu\lambda\zeta} = - 4 !,
\varepsilon_{\mu\nu\lambda\zeta} \varepsilon^{\mu\nu\lambda\rho}
= - 3! \delta^\rho_\zeta$ etc. Here the Greek indices
$\mu, \nu, \lambda....= 0, 1, 2, 3$ correspond to the spacetime directions
on the 4D ordinary manifold and the Latin indices
$i, j, k....= 1, 2, 3$ stand for the space directions only. The 3-vectors
are occasionally represented by the bold faced letters
(i.e. ${\bf B} = B_i, {\bf E} = E_i, {\bf b^{(1)}} = b^{(1)}_i,
{\bf b^{(2)}} = b^{(2)}_i$, etc.)}
ordinary flat Minkowski spacetime manifold
$$
\begin{array}{lcl}
{\cal L}^{(4)}_b &=& -
{\displaystyle \frac{1}{4}}\; F^{\mu\nu} F_{\mu\nu} + B (\partial \cdot A)
+ {\displaystyle \frac{1}{2}}\;
 B^2 - i \partial_\mu \bar C \partial^\mu C, \nonumber\\
&\equiv&
{\displaystyle  \frac{1}{2}}\; ({\bf E}^2 - {\bf B}^2) + B (\partial \cdot A)
+ {\displaystyle \frac{1}{2}}\; B^2 - i \partial_\mu \bar C \partial^\mu C,
\end{array} \eqno(3.1)
$$ where the field strength tensor $F_{\mu\nu} = \partial_\mu
A_\nu - \partial_\nu A_\mu$, constructed from $d = dx^\mu
\partial_\mu$ and the 1-form $A^{(1)} = dx^\mu A_\mu$ through $ F^{(2)} = d
A^{(1)} = \frac{1}{2} (dx^\mu \wedge dx^\nu) F_{\mu\nu}$, has the
electric ($F_{0i} = E_i = {\bf E}$) and the magnetic ($F_{ij} =
\epsilon_{ijk} B_k, B_i = {\bf B} = - \frac{1}{2} \epsilon_{ijk}
F_{jk}$) components and the gauge-fixing term $(\partial \cdot A)
= \partial_0 A_0 - \partial_i A_i$ is constructed by the
application of the nilpotent ($\delta^2 = 0$) co-exterior
derivative $\delta = - * d *$ on the 1-form $A^{(1)} = dx^\mu
A_\mu$ (i.e. $\delta A^{(1)} = (\partial \cdot A)$). Here the
Hodge duality $*$ operation is defined on the 4D Minkowskian flat
spacetime manifold. All the other symbols carry the same meaning
as discussed in  Section 2. The above Lagrangian density can be
linearized by introducing a couple of vector auxiliary fields
${\bf b^{(1)}}, {\bf b^{(2)}}$ as [42] $$
\begin{array}{lcl}
{\cal L}^{(4)}_B &=&  b^{(1)}_i  E_i - \frac{1}{2} ({\bf b^{(1)}})^2
- b^{(2)}_i B_i + \frac{1}{2} ({\bf b^{(2)}})^2 + B (\partial \cdot A)
+ \frac{1}{2} B^2 - i \partial_\mu \bar C \partial^\mu C.
\end{array} \eqno(3.2).
$$
The above Lagrangian density respects the following local, covariant,
continuous, off-shell
nilpotent ($s_{(a)b}^2 = 0$) and anticommuting ($s_b s_{ab} + s_{ab} s_b = 0$)
(anti-)BRST $(s_{(a)b})$ symmetry transformations [42]
$$
\begin{array}{lcl}
s_b A_\mu &=& \partial_\mu C, \qquad s_b C = 0, \qquad
s_b \bar C = i B, \qquad s_b B = 0, \nonumber\\
s_b {\bf B} &=& 0, \qquad s_b {\bf b^{(1)}} = 0, \qquad s_b {\bf b^{(2)}} = 0,
\qquad s_b {\bf E} = 0, \nonumber\\
s_{ab} A_\mu &=& \partial_\mu \bar C, \qquad s_{ab} \bar C = 0, \qquad
s_{ab} C = - i B, \qquad s_{ab} B = 0, \nonumber\\
 s_{ab} {\bf B} &=& 0, \qquad s_{ab} {\bf b^{(1)}} = 0, \qquad
s_{ab} {\bf b^{(2)}} = 0, \qquad
\quad s_{ab} {\bf E} = 0,
\end{array} \eqno(3.3)
$$
because (3.2) transforms to a total derivative under the above
transformations. Furthermore, the same Lagrangian density is endowed with
the following non-local, non-covariant, continuous,
off-shell nilpotent $(s_{(a)d}^2 = 0)$ and anticommuting
($s_d s_{ad} + s_{ad} s_d = 0$) (anti-)co-BRST symmetry transformations
$(s_{(a)d})$ (see, e.g., [40-42] for details)
$$
\begin{array}{lcl}
s_d A_0 &=& i \bar C, \qquad s_d A_i = i
{\displaystyle \frac{\partial_0 \partial_i}{\nabla^2}} \bar C,
\qquad s_d \bar C = 0, \qquad s_d {\bf B} = 0,
\qquad s_d {\bf b^{(1)}} = 0,
\nonumber\\
s_d C &=& +  {\displaystyle \frac{\partial_i b^{(1)}_i} {\nabla^2}},
\;\;\qquad  s_d B = 0, \;\;\qquad s_d {\bf b^{(2)}} = 0, \;\;\qquad
s_d (\partial \cdot A) = 0,
\nonumber\\
s_{ad} A_0 &=& i  C, \qquad s_{ad} A_i = i
{\displaystyle \frac{\partial_0 \partial_i}{\nabla^2}}  C,
\qquad s_{ad} C = 0, \qquad s_{ad} {\bf B} = 0,
\qquad s_{ad} {\bf b^{(1)}} = 0,\nonumber\\
s_{ad} \bar C &=& -  {\displaystyle \frac{\partial_i b^{(1)}_i} {\nabla^2}},
\;\;\qquad  s_{ad} B = 0, \;\;\qquad s_{ad} {\bf b^{(2)}} = 0, \;\;\qquad
s_{ad} (\partial \cdot A) = 0,
\end{array} \eqno(3.4)
$$
where $\nabla^2 = \partial_i \partial_i = (\partial_1)^2
+ (\partial_2)^2 + (\partial_3)^2$.
At this stage, a few comments are in order. (i) It is clear that the
(anti-)BRST symmetry transformations
 are local, covariant, continuous, nilpotent
and anticommuting. In contrast, the (ant-)co-BRST symmetry
transformations are non-local, non-covariant, continuous,
nilpotent and anticommuting. (ii) The nilpotent (anti-)BRST as
well as (anti-)co-BRST transformations keep the magnetic field
${\bf B}$ invariant. (iii) Under the (anti-)BRST and
(anti-)co-BRST transformations, the 2-form $F^{(2)} = d A^{(1)}$
and the 0-form $(\partial\cdot A) = \delta A^{(1)}$ remain
invariant, respectively. (iv) It is evident that the
anticommutator $\{ s_d, s_b \} = \{s_{ab}, s_{ad} \} = s_w$ leads
to the definition of a non-nilpotent bosonic symmetry $s_w$.
However, the exact expressions for these transformations are not
essential for our present discussions. (v) The global scale
transformations on the (anti-)ghost fields define the ghost
symmetry in the theory. The corresponding conserved charge is the
ghost charge $Q_g$. (vi) The above conserved Noether charges
generate the transformations (2.5).\\

\noindent
{\bf 3.1 Superfield formulation of (anti-)BRST symmetries for 4D theory}\\

\noindent
We consider the free four $(3 + 1)$-dimensional
(4D) Abelian gauge theory on a six $(4 + 2)$-dimensional
supermanifold parametrized by the four spacetime $x^\mu (\mu = 0, 1, 2, 3)$
bosonic variables and a couple of odd ($\theta^2 = \bar\theta^2 = 0,
\theta \bar\theta + \bar\theta \theta = 0$)
Grassmannian variables $\theta$ and $\bar\theta$. The local
basic fields $(A_\mu (x), C (x), \bar C (x))$ of the Lagrangian density
(3.1) are now generalized to the superfields
$(B_\mu (x,\theta,\bar\theta), \Phi (x,\theta,\bar\theta),
\bar\Phi (x,\theta,\bar\theta)$ on the six dimensional supermanifold.
These latter superfields can be expanded in terms of the basic fields
as given in (2.7). However, there is a subtle difference between the
expansion on the four $(2 + 2)$-dimensional
(cf. Section 2) and the six $(4 + 2)$-dimensional
supermanifold. For instance, in the following
$$
\begin{array}{lcl}
B_{\mu}\; (x, \theta, \bar \theta) &=& A_{\mu} (x)
+ \;\theta\;  \bar R_\mu (x)
+  \;\bar \theta\;  R_\mu (x)
+ i \;\theta \;\bar \theta \;S_{\mu} (x), \nonumber\\
\Phi\; (x, \theta, \bar \theta) &=& C (x) \;+ i \; \theta\;  \bar B (x)
\;- i \;\bar \theta\;  {\cal B} (x)
\;+ i \;\theta \;\bar \theta \; s (x),
 \nonumber\\
\bar \Phi\; (x, \theta, \bar \theta) &=& \bar C (x)
\;- i \;\theta\;\bar {\cal B} (x)) \;+ i\;\bar \theta\;  B (x)
\;+ i\;\theta\;\bar \theta \; \bar s (x),
\end{array} \eqno(3.5)
$$ the auxiliary scalar fields ${\cal B}$ and $\bar {\cal B}$ are
{\it not} the ones that have been written for the 2D free Abelian
gauge theory. In particular, the auxiliary scalar field ${\cal B}$
appears explicitly in the Lagrangian density (2.2) for the 2D
theory. However, it does not appear explicitly in the Lagrangian
density of the 4D theory. All the rest of the steps are exactly
the same (see, e.g.,  equations (2.8)--(2.12)) as discussed in the
sub-section 2.1 for the discussion of the 2D Abelian theory on a
$(2 + 2)$-dimensional supermanifold. Finally, the horizontality
condition $\tilde d \tilde A^{(1)} = d A^{(1)}$ leads to the
derivation of the nilpotent (anti-)BRST symmetry transformations
(3.3) for the 4D free Abelian gauge theory as expressed below in
the language of the superfield expansion on the six $(4 +
2)$-dimensional supermanifold (see, e.g., [42] for details) $$
\begin{array}{lcl}
B_{\mu}\; (x, \theta, \bar \theta) &=& A_{\mu} (x)
+ \;\theta\; (s_{ab} A_{\mu} (x))
+ \;\bar \theta\; (s_{b} A_{\mu} (x))
+ \;\theta \;\bar \theta \;(s_{b} s_{ab} A_{\mu} (x)), \nonumber\\
\Phi\; (x, \theta, \bar \theta) &=& C (x) \;+ \; \theta\; (s_{ab} C (x))
\;+ \;\bar \theta\; (s_{b} C (x))
\;+ \;\theta \;\bar \theta \;(s_{b}\; s_{ab} C (x)),
 \nonumber\\
\bar \Phi\; (x, \theta, \bar \theta) &=& \bar C (x)
\;+ \;\theta\;(s_{ab} \bar C (x)) \;+\bar \theta\; (s_{b} \bar C (x))
\;+\;\theta\;\bar \theta \;(s_{b} \;s_{ab} \bar C (x)).
\end{array} \eqno(3.6)
$$
The above equation establishes the geometrical interpretation for the
off-shell nilpotent (anti-)BRST charges $Q_{(a)b}$ as the translation
generators $(\partial/\partial\theta)\partial/\partial\bar\theta$ along
the $(\theta)\bar\theta$-directions of the six $(4 + 2)$-dimensional
supermanifold. In fact, the process of translations of the superfields
$(B_\mu, \Phi, \bar \Phi)$ along $(\theta)\bar\theta$-directions of
the supermanifold produces the internal (anti-)BRST symmetry transformations
$s_{(a)b}$ (cf. (3.3)) for the local fields $(A_\mu, C, \bar C)$.\\

\noindent
{\bf 3.2 Hodge duality on $(4 + 2)$-dimensional supermanifold}\\

\noindent To obtain the nilpotent ($s_{(a)d}^2 = 0$) and
anticommuting $(s_d s_{ad} + s_{ad} s_d = 0$) (anti-)co-BRST
transformations $s_{(a)d}$ for the basic fields $(A_\mu, C, \bar
C)$ of the 4D free Abelian gauge theory, we have to exploit the
dual-horizontality condition $\tilde \delta \tilde A^{(1)} =
\delta A^{(1)}$ where $\tilde \delta = - \star \tilde d \star$ and
$\delta = -
* d *$ are the super co-exterior derivative and the ordinary
co-exterior derivative, respectively. These derivatives are
defined  on the six $(4 + 2)$-dimensional supermanifold and the
ordinary 4D Minkowskian spacetime manifold. As discussed earlier,
$\tilde d$ is the super exterior derivative (see, e.g., for the
definition, equation (2.9)) and $\star$ and $*$ are the Hodge
duality operations on the supermanifold and the ordinary manifold,
respectively. The $(4 + 2)$-dimensional supermanifold can support
only three (super) 1-forms as given below $$
\begin{array}{lcl}
{\cal O}_{1} = dx^\mu\; P_\mu, \qquad
{\cal O}_{2} = d \theta\; P_\theta, \qquad
{\cal O}_{3} = d\bar\theta \;P_{\bar\theta}.
\end{array} \eqno(3.7)
$$
However, as will become clear later, a triplet of  (super) 1-forms
can be constructed from the differential $dx^\mu$ that is present in
the definition of ${\cal O}_1$. The Hodge duality $\star$ operation on
the above 1-forms produces the following
5-forms
$$
\begin{array}{lcl}
&& \star\; {\cal O}_1 =  \frac{1}{3!}\;
\varepsilon^{\mu\nu\lambda\zeta}\; (d x_\nu \wedge dx_\lambda \wedge
dx_\zeta \wedge d \theta \wedge d \bar\theta)\; P_\mu , \nonumber\\
&& \star\; {\cal O}_2 = \frac{1}{4!}\; \varepsilon^{\mu\nu\lambda\zeta}\;
(dx_\mu \wedge d x_\nu \wedge dx_\lambda \wedge dx_\zeta
\wedge d \bar\theta)\; P_\theta, \nonumber\\
&& \star\; {\cal O}_3 = \frac{1}{4!}\; \varepsilon^{\mu\nu\lambda\zeta} \;
(dx_\mu \wedge d x_\nu \wedge d x_\lambda \wedge dx_\zeta \wedge d \theta)
\; P_{\bar\theta}.
\end{array} \eqno(3.8)
$$
It will be noted, in the above, that (i) the $\star$ operation acts basically
on the differentials and it does not act on $P$'s. (ii) A linear
combination of the 1-forms of (3.7) can also be considered as 1-form.
(iii) The  double $\star$ operation on the above 1-forms yields
$$
\begin{array}{lcl}
\star\; [\;\star\;{\cal O}_{1}\;] = + \; {\cal O}_1, \qquad
\star\; [\; \star\; {\cal O}_{2}\;]  = - \; {\cal O}_2, \qquad
\star \; [\;\star\; {\cal O}_{3}\;] = - \; {\cal O}_3,
\end{array} \eqno(3.9)
$$
where we have used $\varepsilon^{\mu\nu\lambda\zeta}
\varepsilon_{\nu\lambda\zeta\rho} = + 3! \delta^\mu_\rho,
\varepsilon^{\mu\nu\lambda\zeta}
\varepsilon_{\mu\nu\lambda\zeta} = - 4!$. Furthermore,
we have used the $\star$ operation on the 5-forms which are
found to be dual to 1-forms. In fact, the six $(4 + 2)$-dimensional
supermanifold can support five independent 5-forms:
$$
\begin{array}{lcl}
&&\tilde \phi_{1} = \frac{1}{3!}\;
(dx^\mu \wedge dx^\nu \wedge dx^\lambda
\wedge d \theta \wedge d \bar\theta)\;
\tilde {\cal F}_{\mu\nu\lambda\theta\bar\theta}, \nonumber\\
&&\tilde \phi_{2} = \frac{1}{3!}\;
(dx^\mu \wedge dx^\nu \wedge dx^\lambda  \wedge
d \theta \wedge d \theta)\;
\tilde {\cal F}_{\mu\nu\lambda\theta\theta}, \nonumber\\
&&\tilde \phi_{3} = \frac{1}{3!}\;
(dx^\mu \wedge dx^\nu \wedge dx^\lambda
\wedge d \bar\theta \wedge d \bar\theta)\;
\tilde {\cal F}_{\mu\nu\lambda\bar\theta\bar\theta},\nonumber\\
&& \tilde \phi_4 = \frac{1}{4!}\;
(dx^\mu \wedge dx^\nu \wedge dx^\lambda  \wedge dx^\zeta
\wedge d \theta)\;
\tilde {\cal F}_{\mu\nu\lambda\zeta\theta}, \nonumber\\
&& \tilde \phi_5 = \frac{1}{4!}\;
(dx^\mu \wedge dx^\nu \wedge dx^\lambda  \wedge dx^\zeta
\wedge d \bar\theta)\;
\tilde {\cal F}_{\mu\nu\lambda\zeta\bar\theta}.
\end{array} \eqno(3.10)
$$
The Hodge duality $\star$ operation on the wedge products of the
differentials, present in the above 5-forms,
yields the following 1-form differentials
\footnote{It will be noted that there are
three 1-form differentials $ (dx_\mu), s^{\theta\theta} (dx_\mu),
s^{\bar\theta\bar\theta} (dx_\mu)$, constructed from $(dx_\mu)$,
because the dual 5-form differentials on supermanifold
are different for each individual of them. For
instance, $\star\; (dx_\mu)
= \frac{1}{3!} \varepsilon_{\mu\nu\lambda\zeta}
(dx^\nu \wedge dx^\lambda \wedge dx^\zeta \wedge d\theta \wedge
d \bar\theta), \;
\star\; [ s^{\theta\theta} (dx_\mu) ]
= \frac{1}{3!} \varepsilon_{\mu\nu\lambda\zeta}
(dx^\nu \wedge dx^\lambda \wedge dx^\zeta \wedge d\theta \wedge d\theta),
\;\star\; [s^{\bar\theta\bar\theta} (dx_\mu) ]
= \frac{1}{3!} \varepsilon_{\mu\nu\lambda\zeta} (dx^\nu \wedge dx^\lambda
\wedge dx^\zeta \wedge d\bar\theta \wedge d\bar\theta)$. Only for the sake
of brevity, a single 1-form $dx^\mu P_\mu$
(constructed from $dx^\mu$) is given in (3.7). It will be noted that such kind
of a triplet of superforms cannot be associated with ${\cal O}_2$ and
${\cal O}_3$ because their Hodge duals are not well-defined on the
six $(4 + 2)$-dimensional supermanifold of our present discussion.}
$$
\begin{array}{lcl}
&&\star\;
(dx^\mu \wedge dx^\nu \wedge dx^\lambda
\wedge d \theta \wedge d \bar\theta)\;
=  \varepsilon^{\mu\nu\lambda\zeta}\; (dx_\zeta), \nonumber\\
&&\star\;
(dx^\mu \wedge dx^\nu \wedge dx^\lambda  \wedge
d \theta \wedge d \theta)\;
= s^{\theta\theta}\;
\varepsilon^{\mu\nu\lambda\zeta}\; (dx_\zeta), \nonumber\\
&&\star\;
(dx^\mu \wedge dx^\nu \wedge dx^\lambda
\wedge d \bar\theta \wedge d \bar\theta)\;
= s^{\bar\theta\bar\theta}\;
\varepsilon^{\mu\nu\lambda\zeta}\; (dx_\zeta), \nonumber\\
&& \star\;
(dx^\mu \wedge dx^\nu \wedge dx^\lambda  \wedge dx^\zeta
\wedge d \theta)\;
= \varepsilon^{\mu\nu\lambda\zeta}\; (d\bar\theta), \nonumber\\
&& \star\;
(dx^\mu \wedge dx^\nu \wedge dx^\lambda  \wedge dx^\zeta
\wedge d \bar\theta)\;
= \varepsilon^{\mu\nu\lambda\zeta}\; (d\bar\theta).
\end{array} \eqno(3.11)
$$
The presence of the symmetric constants, $s^{\theta\theta}$ and
$s^{\bar\theta\bar\theta}$ on the r.h.s. of (3.11), enforces the following
Hodge duality $\star$ operation
$$
\begin{array}{lcl}
&&\star\; [\;s^{\theta\theta}\; (dx^\mu)\;]
= \frac{1}{3!}\;\varepsilon^{\mu\nu\lambda\zeta}\;
(dx_\nu\wedge dx_\lambda\wedge dx_\zeta \wedge d\theta \wedge d\theta),
\nonumber\\
&&\star\; [\;s^{\bar\theta\bar\theta}\; (dx^\mu)\;]
= \frac{1}{3!}\;\varepsilon^{\mu\nu\lambda\zeta}\;
(dx_\nu\wedge dx_\lambda\wedge dx_\zeta
\wedge d\bar\theta \wedge d\bar\theta).
\end{array} \eqno(3.12)
$$
Taking into account (3.8), (3.11) and (3.12), it is clear that the double
$\star$ operation on the 5-forms in (3.10) leads to
$$
\begin{array}{lcl}
&&\star\;[\;\star\;\tilde \phi_{1}\;] = + \; \tilde \phi_1, \qquad
\star\;[\; \star\;\tilde \phi_{2}\;] = +\; \tilde \phi_2,
\qquad \star\; [\; \star\; \tilde \phi_{3}\;] = +\; \tilde \phi_3,
\nonumber\\
&& \star\;[\;\star\;\tilde \phi_4 \;] = \; - \; \tilde \phi_4, \;\;\qquad\;\;
\star\; [\; \star\; \tilde \phi_5 \;] = \; -\; \tilde \phi_5.
\end{array} \eqno(3.13)
$$ The six $(4 + 2)$-dimensional supermanifold can support six
2-forms analogous to (2.24). Their explicit expressions are as
under $$
\begin{array}{lcl}
&& \tilde \chi_{1} = \frac{1}{2!}\; (dx^\mu \wedge dx^\nu) \;
\tilde S_{\mu\nu}, \;\qquad\;
\tilde \chi_{2} = (d \theta \wedge d \bar\theta)\;
\tilde S_{\theta\bar\theta}, \nonumber\\
&&
\tilde \chi_{3} = (d\theta \wedge d \theta) \;
\tilde S_{\theta\theta}, \;\;\;\;\;\qquad\;\;\;\;
\tilde \chi_{4} =
(d\bar\theta \wedge d \bar\theta)\;
\tilde S_{\bar\theta\bar\theta}, \nonumber\\
&&\tilde \chi_{5} = (dx^\mu \wedge d\theta)
\;\tilde S_{\mu\theta}, \;\;\;\qquad \;\;\;\;
\tilde \chi_{6} = (dx^\mu \wedge d\bar\theta)
\;\tilde S_{\mu\bar\theta}.
\end{array} \eqno(3.14)
$$
It is clear that the components
$\tilde S_{\mu\nu}, \tilde S_{\mu\theta}, \tilde S_{\mu\bar\theta}$ are
 antisymmetric. However, the components with the Grassmannian indices
$\tilde S_{\theta\theta}, \tilde S_{\bar\theta\bar\theta},
\tilde S_{\theta\bar\theta}$ are symmetric.
On the above supermanifold, the operation of a single Hodge duality
$\star$ operation leads to the definition of 4-forms which are dual
to the above 2-forms. In fact, a single $\star$ operation on the
wedge products of the differentials of the above 2-forms, are
$$
\begin{array}{lcl}
&& \star\; (dx^\mu \wedge dx^\nu) = \frac{1}{2!}
\;\varepsilon^{\mu\nu\lambda\zeta}\;
(dx_\lambda \wedge dx_\zeta \wedge d \theta \wedge d \bar\theta), \nonumber\\
&& \star\; (d \theta \wedge d\bar\theta)
= \frac{1}{4!} \varepsilon^{\mu\nu\lambda\zeta}
(dx_\mu \wedge dx_\nu\wedge dx_\lambda \wedge dx_\zeta), \nonumber\\
&& \star\; (d \theta \wedge d\theta) = \frac{1}{4!} s^{\theta\theta}
\varepsilon^{\mu\nu\lambda\zeta} (dx_\mu \wedge dx_\nu \wedge dx_\lambda
\wedge dx_\zeta), \nonumber\\
&& \star\; (d \bar\theta \wedge d\bar\theta)
= \frac{1}{4!} s^{\bar\theta\bar\theta}
\varepsilon^{\mu\nu\lambda\zeta} (dx_\mu \wedge dx_\nu
\wedge dx_\lambda \wedge dx_\zeta), \nonumber\\
&& \star\;(dx^\mu \wedge d \theta) =
\frac{1}{3!} \varepsilon^{\mu\nu\lambda\zeta}\; (dx_\nu \wedge dx_\lambda
\wedge dx_\zeta \wedge d \bar\theta), \nonumber\\
&& \star\;(dx^\mu \wedge d \bar \theta) =
\frac{1}{3!} \varepsilon^{\mu\nu\lambda\zeta}\; (dx_\nu \wedge dx_\lambda
\wedge dx_\zeta \wedge d \theta).
\end{array} \eqno(3.15)
$$
This shows that the wedge products of the differentials
corresponding to the 4-forms in the above equations are Hodge dual to
the wedge products of the differentials corresponding to 2-forms considered
(cf. (3.14)) on the six $(4 + 2)$-dimensional supermanifold. The total
number of the independent 4-forms on the above supermanifold are
$$
\begin{array}{lcl}
&& \tilde \tau_{1} = \frac{1}{2!} (dx^\mu \wedge dx^\nu \wedge d
\theta \wedge d \bar\theta) \tilde T_{\mu\nu\theta\bar\theta},
\qquad \tilde \tau_{2} = \frac{1}{2!} (dx^\mu \wedge dx^\nu \wedge
d \bar\theta \wedge d \bar\theta) \tilde
T_{\mu\nu\bar\theta\bar\theta}, \nonumber\\ && \tilde \tau_{3} =
\frac{1}{2!}
 (dx^\mu \wedge dx^\nu \wedge d\theta \wedge d \theta) \;
\tilde T_{\mu\nu\theta\theta}, \qquad
\tilde \tau_{4} = \frac{1}{3!}
(dx^\mu \wedge d x^\nu \wedge dx^\lambda \wedge d \bar\theta)\;
\tilde T_{\mu\nu\lambda\bar\theta}, \nonumber\\
&&\tilde \tau_{5} = \frac{1}{3!}
 (dx^\mu \wedge dx^\nu \wedge dx^\lambda \wedge d\theta)
\;\tilde T_{\mu\nu\lambda\theta}, \quad\;
\tilde \tau_6 = \frac{1}{4!}
 (dx^\mu \wedge dx^\nu \wedge dx^\lambda \wedge dx^\zeta)
\; \tilde T_{\mu\nu\lambda\zeta}.
\end{array} \eqno(3.16)
$$
It will be noted that the 4-forms with the wedge products
$(dx_\mu \wedge dx_\nu \wedge dx_\lambda\wedge dx_\zeta),
(dx_\mu \wedge dx_\nu \wedge dx_\lambda\wedge dx_\zeta)\; s^{\theta\theta},
(dx_\mu \wedge dx_\nu \wedge dx_\lambda\wedge dx_\zeta)\;
s^{\bar\theta\bar\theta}$ are different because their dual 2-forms are
different as can be seen from (3.15). However, for the sake of brevity,
we have chosen only one
\footnote{In principle, one can define a triplet of $\tilde \tau_6$ form
in (3.16). These are $
\tilde \tau_6^{(1)}
= \frac{1}{4!} (dx^\mu \wedge dx^\nu \wedge dx^\lambda \wedge dx^\zeta)
\; \tilde T^{(1)}_{\mu\nu\lambda\zeta},
\tilde \tau_6^{(2)} = \frac{1}{4!}
 (dx^\mu \wedge dx^\nu \wedge dx^\lambda \wedge dx^\zeta)\;
s^{\theta\theta}\;
\; \tilde T^{(2)}_{\mu\nu\lambda\zeta},
\tilde \tau_6^{(3)} = \frac{1}{4!}
(dx^\mu \wedge dx^\nu \wedge dx^\lambda \wedge dx^\zeta)\;
s^{\bar\theta\bar\theta}\;
\; \tilde T^{(3)}_{\mu\nu\lambda\zeta}$. It is obvious that the Hodge dual
of these forms are distinct and different. For obvious reasons,
no other forms in (3.16) support
such kind of a triplet of superforms (as their Hodge dual forms are not
well-defined on the supermanifold in the sense that they will contain more
than two Grassmannian wedge products).}
of these in (3.16).
A single $\star$ operation on the wedge products of the differentials
corresponding to 4-forms are
$$
\begin{array}{lcl}
&& \star\; (dx^\mu \wedge dx^\nu \wedge d \theta \wedge d
\bar\theta) = \frac{1}{2!} \varepsilon^{\mu\nu\lambda\zeta}\;
(dx_\lambda \wedge d x_\zeta), \nonumber\\ &&\star\; (dx^\mu
\wedge dx^\nu \wedge d \bar\theta \wedge d \bar\theta)\; =
s^{\bar\theta\bar\theta}\; \frac{1}{2!}\;
\varepsilon^{\mu\nu\lambda\zeta} (dx_\lambda \wedge dx_\zeta),
\nonumber\\ && \star\;(dx^\mu \wedge dx^\nu \wedge d\theta \wedge
d \theta) \; = s^{\theta\theta} \frac{1}{2!}
\varepsilon^{\mu\nu\lambda\zeta} (dx_\lambda \wedge dx_\zeta),
\nonumber\\ &&\star \;(dx^\mu \wedge d x^\nu \wedge dx^\lambda
\wedge d \bar\theta)\; = \varepsilon^{\mu\nu\lambda\zeta}
(dx_\zeta \wedge d \theta), \nonumber\\ &&\star\; (dx^\mu \wedge
dx^\nu \wedge dx^\lambda \wedge d\theta) =
\varepsilon^{\mu\nu\lambda\zeta} (dx_\zeta \wedge d \bar\theta),
\nonumber\\ &&\star\;(dx^\mu \wedge dx^\nu \wedge dx^\lambda
\wedge dx^\zeta) = \varepsilon^{\mu\nu\lambda\zeta} \; (d\theta
\wedge d\bar\theta).
\end{array} \eqno(3.17)
$$
It is clear from (3.15)--(3.17) that one can compute now the double
$\star$ operations on the 2-forms as well as 4-forms. Finally, we focus
on the independent 3-forms that can be supported on the $(4 + 2)$-dimensional
supermanifold. There exist six such forms:
$$
\begin{array}{lcl}
&& \sigma_{1} = \frac{1}{2!} (dx^\mu \wedge dx^\nu \wedge d \theta) \;
R_{\mu\nu\theta}, \;\;\;\qquad\;\;\;
\sigma_{2} = \frac{1}{2!} (dx^\mu \wedge dx^\nu \wedge d \bar\theta)\;
R_{\mu\nu\bar\theta}, \nonumber\\
&&
\sigma_{3} = (dx^\mu \wedge d\theta \wedge d \theta) \;
R_{\mu\theta\theta}, \;\;\;\;\qquad\;\;\;\;\;
\sigma_{4} =
(dx^\mu \wedge d\bar\theta \wedge d \bar\theta)\;
R_{\mu\bar\theta\bar\theta}, \nonumber\\
&&\sigma_{5} = (dx^\mu \wedge d\theta \wedge d \bar\theta)
\;R_{\mu\theta\bar\theta}, \;\;\;\;\qquad\;\;\;\;
 \sigma_{6} = \frac{1}{3!} (dx^\mu \wedge dx^\nu \wedge d x^\lambda) \;
R_{\mu\nu\lambda}.
\end{array} \eqno(3.18)
$$
A single Hodge duality $\star$ operation on the above 3-forms will
lead to the derivation of the dual 3-forms on the six $(4 + 2)$-dimensional
supermanifold. Such an operation will affect the wedge products of the
differentials as given below
$$
\begin{array}{lcl}
&& \star\; (dx^\mu \wedge dx^\nu \wedge d \theta) =
\frac{1}{2!}\;\varepsilon^{\mu\nu\lambda\zeta}\;
(dx_\lambda \wedge dx_\zeta \wedge d \bar\theta), \nonumber\\
&&\star\; (dx^\mu \wedge dx^\nu \wedge d \bar\theta) =
\frac{1}{2!}\;\varepsilon^{\mu\nu\lambda\zeta}\;
(dx_\lambda \wedge dx_\zeta \wedge d \theta), \nonumber\\
&& \star\; (dx^\mu \wedge d \theta \wedge d \bar\theta)
= \frac{1}{3!}\;\varepsilon^{\mu\nu\lambda\zeta}\;
(d x_\nu \wedge dx_\lambda \wedge dx_\zeta), \nonumber\\
&&\star\; (dx^\mu \wedge d \theta \wedge d \theta)
= \frac{1}{3!}\;\varepsilon^{\mu\nu\lambda\zeta}\;
(d x_\nu \wedge dx_\lambda \wedge d x_\zeta)\; s^{\theta\theta}, \nonumber\\
&& \star\; (dx^\mu \wedge d \bar \theta \wedge d \bar \theta)
= \frac{1}{3!}\;\varepsilon^{\mu\nu\lambda\zeta}\;
(d x_\nu \wedge dx_\lambda \wedge dx_\zeta)\; s^{\bar\theta \bar\theta},
\nonumber\\
&& \star\; (dx^\mu \wedge dx^\nu \wedge d x^\lambda) =
\;\varepsilon^{\mu\nu\lambda\zeta}\;
(dx_\zeta \wedge d \theta \wedge d \bar\theta).
\end{array} \eqno(3.19)
$$
As expected, there are three 3-forms constructed by the wedge products
of the spacetime differentials
$(dx^\mu \wedge dx^\nu \wedge dx^\lambda),
(dx^\mu \wedge dx^\nu \wedge dx^\lambda) s^{\theta\theta},
(dx^\mu \wedge dx^\nu \wedge dx^\lambda) s^{\bar\theta\bar\theta}$ whose
Hodge duals are different 3-forms as given below
$$
\begin{array}{lcl}
&&\star \; (dx^\mu \wedge dx^\nu \wedge dx^\lambda)
= \varepsilon^{\mu\nu\lambda\zeta}\;
(dx_\zeta \wedge d\theta \wedge d \bar\theta), \nonumber\\
&&\star \; [\; (dx^\mu \wedge dx^\nu \wedge dx^\lambda)\; s^{\theta\theta}\; ]
= \varepsilon^{\mu\nu\lambda\zeta}\;
(dx_\zeta \wedge d\theta \wedge d \theta), \nonumber\\
&&\star \; [\; (dx^\mu \wedge dx^\nu \wedge dx^\lambda)\;
s^{\bar\theta\bar\theta} \;]
= \varepsilon^{\mu\nu\lambda\zeta}\;
(dx_\zeta \wedge d\bar\theta \wedge d \bar\theta).
\end{array} \eqno(3.20)
$$
The above considerations allow us to define the following triplet
of $\sigma_6$ of (3.18)
$$
\begin{array}{lcl}
&& \sigma_6^{(1)} = \frac{1}{3!} (dx^\mu \wedge dx^\nu \wedge dx^\lambda)\;
R^{(1)}_{\mu\nu\lambda}, \qquad\;
\sigma_6^{(2)} = \frac{1}{3!}
 (dx^\mu \wedge dx^\nu \wedge dx^\lambda)\;s^{\theta\theta}
\;R^{(2)}_{\mu\nu\lambda}, \nonumber\\
&& \sigma_6^{(3)} = \frac{1}{3!}
 (dx^\mu \wedge dx^\nu \wedge dx^\lambda)\;
s^{\bar\theta\bar\theta}\;
R^{(3)}_{\mu\nu\lambda}.
\end{array} \eqno(3.21)
$$
However, for the sake of brevity, we have taken only one of the above
triplets in equation (3.18).
It is now straightforward to check that the double Hodge duality
$\star$ operations on the 3-forms of (3.18) yield the following
$$
\begin{array}{lcl}
&& \star\; [\; \star \;\sigma_{1}\;] = \;- \; \sigma_1,
\;\;\;\qquad\;\;\;
\star\; [\; \star\;\sigma_{2}\;]
= \;-\; \sigma_2, \nonumber\\
&&
\star\; [\; \star\; \sigma_{3}\;] = \;+\; \sigma_3,
\;\;\;\qquad\;\;\;
\star \; [\; \star\; \sigma_{4}\;] = \; +\; \sigma_4,
\nonumber\\
&&
\star\; [\; \star\; \sigma_{5}\;]
=\; +\; \sigma_5, \;\;\;\qquad\;\;\;
\star\; [\; \star\; \sigma_{6}\;] =\; + \; \sigma_6.
\end{array} \eqno(3.22)
$$
Finally, we do know that the six $(4 + 2)$-dimensional supermanifold
can support three 6-forms, modulo some constant factors, as given below
$$
\begin{array}{lcl}
&&\tilde \Psi_{1} = \frac{1}{4!}\;
(dx^\mu \wedge dx^\nu \wedge dx^\lambda \wedge dx^\zeta
\wedge d \theta \wedge d \bar\theta)\;
\tilde {\cal G}_{\mu\nu\lambda\zeta\theta\bar\theta}, \nonumber\\
&&\tilde \Psi_{2} = \frac{1}{4!}\;
(dx^\mu \wedge dx^\nu \wedge dx^\lambda  \wedge d x^\zeta \wedge
d \theta \wedge d \theta)\;
\tilde {\cal G}_{\mu\nu\lambda\zeta\theta\theta}, \nonumber\\
&&\tilde \Psi_{3} = \frac{1}{4!}\;
(dx^\mu \wedge dx^\nu \wedge dx^\lambda \wedge dx^\zeta
\wedge d \bar\theta \wedge d \bar\theta)\;
\tilde {\cal G}_{\mu\nu\lambda\zeta\bar\theta\bar\theta}.
\end{array} \eqno(3.23)
$$
A single Hodge duality $\star$ operation on the above 6-forms
produces 0-form scalars on the six dimensional supermanifold. Such
an operation on the wedge products of the differentials are
$$
\begin{array}{lcl}
&&\star\;
(dx^\mu \wedge dx^\nu \wedge dx^\lambda \wedge dx^\zeta
\wedge d \theta \wedge d \bar\theta)\;
= \varepsilon^{\mu\nu\lambda\zeta}, \nonumber \\
&&\star\;
(dx^\mu \wedge dx^\nu \wedge dx^\lambda  \wedge d x^\zeta \wedge
d \theta \wedge d \theta)\;
= \varepsilon^{\mu\nu\lambda\zeta}\; s^{\theta\theta}, \nonumber\\
&&\star\;
(dx^\mu \wedge dx^\nu \wedge dx^\lambda \wedge dx^\zeta
\wedge d \bar\theta \wedge d \bar\theta)\;
= \varepsilon^{\mu\nu\lambda\zeta} \;s^{\bar\theta\bar\theta}.
\end{array} \eqno(3.24)
$$
Two consecutive $\star$ operation on the 6-forms of (3.23) leads to
$$
\begin{array}{lcl}
\star\; [\; \star\; \Psi_{1}\;]  = -\; \Psi_1, \qquad
\star\; [\; \star\; \Psi_{2}\;]  = -\; \Psi_2, \qquad
\star\; [\; \star\; \Psi_{3}\;]  = -\; \Psi_3.
\end{array} \eqno(3.25)
$$
It is evident that we have collected, in the present section, all
the possible super-forms, their single Hodge dual- as well as
their double Hodge dual superforms, etc., that could be defined on
the $(4 + 2)$-dimensional supermanifold.\\

\noindent
{\bf 3.3 Superfield formulation of (anti-)co-BRST symmetries for 4D theory}\\

\noindent As evident from (3.4) that the non-local, non-covariant,
continuous, off-shell nilpotent and anticommuting ($s_d s_{ad} +
s_{ad} s_d = 0$) (anti-)co-BRST symmetries $s_{(a)d}$ do exist for
the 4D free Abelian gauge theory. To obtain these symmetries in
the framework of superfield formulation, we have to exploit the
dual-horizontality condition $\tilde \delta \tilde A^{(1)} =
\delta A^{(1)}$ on the six $(4 + 2)$-dimensional supermanifold. It
is clear that the r.h.s. of the above condition (i.e $\delta
A^{(1)} = - * d * A^{(1)} = (\partial \cdot A)$) is the usual
gauge-fixing term on the ordinary 4D spacetime manifold. For the
computation of the l.h.s. $\tilde \delta \tilde A^{(1)} = - \star
\delta \star \tilde A^{(1)}$, we first concentrate on the dual
$(\star \tilde A^{(1)} = \star dZ^M \tilde A_M)$ of the super
1-form connection $\tilde A^{(1)}$. The ensuing expression for
$(\star \tilde A^{(1)})$, due to the Hodge duality operation given
in (3.8) and definition (2.9), is $$
\begin{array}{lcl}
\star\; \tilde A^{(1)} &=&
\frac{1}{3!}\;\varepsilon^{\mu\nu\lambda\zeta} \; (dx_\nu \wedge
dx_\lambda \wedge dx_\zeta \wedge d \theta \wedge d \bar\theta)\;
B_\mu (x,\theta,\bar\theta)\nonumber\\ &+& \frac{1}{4!}\;
\varepsilon_{\mu\nu\lambda\zeta}\; (dx^\mu \wedge dx^\nu \wedge
dx^\lambda \wedge dx^\zeta \wedge d \bar\theta)\; \bar \Phi
(x,\theta, \bar\theta) \nonumber\\ &+& \frac{1}{4!}\;
\varepsilon_{\mu\nu\lambda\zeta}\; (dx^\mu \wedge dx^\nu \wedge
dx^\lambda \wedge dx^\zeta \wedge d \theta)\;  \Phi
(x,\theta,\bar\theta),
\end{array} \eqno(3.26)
$$
which is nothing but the 5-form defined on the six $(4 + 2)$-dimensional
supermanifold. Applying now the super exterior derivative $\tilde d
= dZ^M \partial_M$ on the above 5-form, we obtain the following 6-form
$$
\begin{array}{lcl}
\tilde d\; (\star \tilde A^{(1)}) &=& \frac{1}{3!}
\varepsilon^{\mu\nu\lambda\zeta}\; (dx_\rho \wedge dx_\nu \wedge
dx_\lambda \wedge d x_\zeta \wedge d\theta \wedge d \bar\theta)\;
(\partial^\rho B_\mu) (x,\theta,\bar\theta) \nonumber\\ &-&
\frac{1}{4!} \varepsilon_{\mu\nu\lambda\zeta}\; (dx^\mu \wedge
dx^\nu \wedge dx^\lambda \wedge dx^\zeta \wedge d\theta \wedge
d\bar\theta)\; (\partial_\theta \bar \Phi) (x,\theta,\bar\theta)
\nonumber\\ &-&
 \frac{1}{4!} \varepsilon_{\mu\nu\lambda\zeta}\;
(dx^\mu \wedge dx^\nu \wedge dx^\lambda \wedge dx^\zeta
 \wedge d \bar\theta \wedge d\bar\theta)\;
(\partial_{\bar\theta} \bar \Phi) (x,\theta,\bar\theta) \nonumber\\
&-& \frac{1}{4!} \varepsilon_{\mu\nu\lambda\zeta}\;
(dx^\mu \wedge dx^\nu \wedge dx^\lambda \wedge dx^\zeta
\wedge d\theta \wedge d \bar \theta)\;
(\partial_{\bar\theta} \Phi) (x,\theta,\bar\theta) \nonumber\\
&-& \frac{1}{4!} \varepsilon_{\mu\nu\lambda\zeta}\;
(dx^\mu \wedge dx^\nu \wedge dx^\lambda \wedge dx^\zeta
 \wedge d\theta \wedge d \theta)\;
(\partial_\theta \Phi) (x,\theta,\bar\theta).
\end{array} \eqno(3.27)
$$
It should be noted that all the wedge products with more than four
spacetime differentials and two Grassmannian differentials have been
 dropped out
because on a $(4 + 2$)-dimensional supermanifold one cannot define
such kind of differential forms. On (3.27), we now apply another $(- \star)$
to obtain a super 0-form (superscalar) by exploiting the Hodge duality
operation defined in (3.24). Such a superscalar is
$$
\begin{array}{lcl}
\tilde \delta \tilde A^{(1)} = - \star \; \tilde d \; \star \;
\tilde A^{(1)} = (\partial \cdot B) - (\partial_\theta \bar\Phi +
\partial_{\bar \theta} \Phi) - s^{\theta\theta}\; (\partial_\theta
\Phi) - s^{\bar\theta\bar\theta}\; (\partial_{\bar\theta} \bar
\Phi).
\end{array} \eqno(3.28)
$$ Equating the above superscalar with the ordinary scalar $\delta
A^{(1)}$, due to the dual-horizontality condition ($\tilde \delta
\tilde A^{(1)} = \delta A^{(1)})$, we obtain the following
relationships $$
\begin{array}{lcl}
(\partial \cdot B) - (\partial_\theta \bar \Phi +
\partial_{\bar\theta} \Phi) = (\partial \cdot A), \qquad
\partial_\theta \Phi = 0, \qquad \partial_{\bar\theta} \bar \Phi = 0.
\end{array} \eqno(3.29)
$$
The insertion of the most general super expansions (cf. (3.5)) leads
to the following restrictions on the secondary fields of expansion (3.5):
$$
\begin{array}{lcl}
&&(\partial \cdot R) (x) = 0, \;\;\qquad\; (\partial \cdot \bar R) (x) = 0,
\;\;\qquad\;
(\partial \cdot S) (x) = 0, \nonumber\\
&& s (x) = \bar s (x) = B (x) = \bar B (x) = 0, \;\;\;\qquad \;\;
{\cal B} (x) + \bar {\cal B} (x) = 0.
\end{array} \eqno(3.30)
$$
Consistent with the statements made after (3.5), the following choices of
the secondary fields in terms of the basic fields (see, e.g., [42]
for details)
$$
\begin{array}{lcl}
&& R_0 = i \bar C, \qquad
R_i = i {\displaystyle
\frac{\partial_0\partial_i}{\nabla^2}} \bar C,
\qquad \bar R_0 = i C,
\qquad
 \bar R_i = i {\displaystyle
\frac{\partial_0\partial_i}{\nabla^2}} C, \nonumber\\
&&{\cal B} = + i {\displaystyle
\frac{\partial_i b^{(1)}_i}{\nabla^2}},  \qquad
\bar {\cal B} = - i {\displaystyle
\frac{\partial_i b^{(1)}_i}{\nabla^2}}, \qquad
S_0 = {\displaystyle \frac{\partial_i b^{(1)}_i}{\nabla^2}},  \qquad
S_i =  {\displaystyle \frac{\partial_0\partial_i}{\nabla^2}}\;\Bigl (
{\displaystyle \frac{\partial_j b^{(1)}_j}{\nabla^2}} \Bigr ),
\end{array} \eqno(3.31)
$$
do satisfy all the above conditions (3.30), emerging from the application
of the dual-horizontality condition. It is worth emphasizing, at this juncture,
that the auxiliary field ${\bf b^{(2)}}$ has {\it not} been taken into account
in the expansion (3.5) as well as in the choices (3.31) because this field
(and its equivalent magnetic field ${\bf B}$) do not appear in any
transformations listed in (3.3) and (3.4). Furthermore, this field, on its
own, does not transform under (co-)BRST transformations. In terms of the
transformations in (3.4) and expressions (3.31), we obtain the following
expansions
$$
\begin{array}{lcl}
B_{0}\; (x, \theta, \bar \theta) &=& A_{0} (x)
+ \;\theta\; (s_{ad} A_{0} (x))
+ \;\bar \theta\; (s_{d} A_{0} (x))
+ \;\theta \;\bar \theta \;(s_{d} s_{ad} A_{0} (x)), \nonumber\\
B_{i}\; (x, \theta, \bar \theta) &=& A_{i} (x)
+ \;\theta\; (s_{ad} A_{i} (x))
+ \;\bar \theta\; (s_{d} A_{i} (x))
+ \;\theta \;\bar \theta \;(s_{d} s_{ad} A_{i} (x)), \nonumber\\
\Phi\; (x, \theta, \bar \theta) &=& C (x) \;+ \; \theta\; (s_{ad} C (x))
\;+ \;\bar \theta\; (s_{d} C (x))
\;+ \;\theta \;\bar \theta \;(s_{d}\; s_{ad} C (x)),
 \nonumber\\
\bar \Phi\; (x, \theta, \bar \theta) &=& \bar C (x)
\;+ \;\theta\;(s_{ad} \bar C (x)) \;+ \;\bar \theta\; (s_{d} \bar C (x))
\;+\;\theta\;\bar \theta \;(s_{d} \;s_{ad} \bar C (x)).
\end{array} \eqno(3.32)
$$
The above expansion does establish the geometrical interpretation for
the conserved and nilpotent (anti-)co-BRST charges $Q_{(a)d}$ as the
translation generators along the Grassmannian directions of the
six $(4 + 2)$-dimensional supermanifold. In fact, there exists some
inter-connections among the nilpotent transformations $s_{(a)d}$,
the translations generators along the Grassmannian directions of the
supermanifold and the nilpotent charges $Q_{(a)d}$, as
$$
\begin{array}{lcl}
s_{d} \leftrightarrow \mbox{Lim}_{\theta \to 0}
{\displaystyle \frac{\partial}{\partial \bar\theta}} \leftrightarrow Q_d,
\qquad
s_{ad} \leftrightarrow \mbox{Lim}_{\bar \theta \to 0}
{\displaystyle \frac{\partial}{\partial \theta}} \leftrightarrow Q_{ad}.
\end{array} \eqno(3.33)
$$
The above relationship is the analogue of exactly the same kind of
relation existing in the context of the nilpotent (anti-)BRST symmetries
(cf. (2.14)).\\

\noindent
{\bf 4 Conclusions}\\

\noindent In our present investigation, we have been able to
define a consistent Hodge duality $\star$ operation on (i) the
four $(2 + 2)$-dimensional supermanifold, and (ii) the six $(4 +
2)$-dimensional supermanifold. These definitions are essential for
the derivation of the nilpotent ($s_{(a)d}^2 = 0$) (anti-)co-BRST
symmetries $s_{(a)d}$ for (i) the two $(1 + 1)$-dimensional (2D)
free Abelian gauge theory, and (ii) the four $(3 + 1)$-dimensional
(4D) free Abelian gauge theory in the framework of superfield
formulation. In fact, the above 2D- and 4D free Abelian gauge
theories (described by the local fields that take values on the 2D
and 4D flat Minkowskian spacetime manifold) are considered on the
four $(2 + 2)$-dimensional- and six $(4 + 2)$-dimensional
supermanifolds, respectively. Our study on these supermanifolds
(described by the superfields that take values on the
supermanifold parametrized by the superspace variables $Z^M=
(x^\mu, \theta, \bar\theta)$) does provide the geometrical origin
and interpretation for the nilpotent (anti-)BRST- and
(anti-)co-BRST symmetries (and the corresponding nilpotent
generators). The physical application of a consistent definition
of the Hodge duality $\star$ operation turns up in the context of
the dual-horizontality condition $\tilde \delta \tilde A^{(1)} =
\delta A^{(1)}$ where the use of the super co-exterior derivative
$\tilde \delta = - \star \tilde d \star$ (on the l.h.s.) does
require a consistent definition of the $\star$ operation. In fact,
the existence of the nilpotent (anti-)co-BRST symmetry
transformations owes its origin to the (super) co-exterior
derivatives where the definition of $\star$ plays a very decisive
role. In the language of physics, it is the gauge-fixing term of
the (anti-)BRST invariant Lagrangian density of a gauge theory
that remains invariant under the (anti-)co-BRST transformations
(cf. Sections 2 and 3). This statement has been shown to be true
for both the 2D- and 4D free Abelian gauge theories where there is
no interaction between the $U(1)$ gauge field and the matter
fields.

One of the novel and the most decisive ingredients in our whole discussion
is the introduction of the constant symmetric parameters $s^{\theta\theta}$
and $s^{\bar\theta\bar\theta}$ in the definition of the Hodge duality
$\star$ operation on the wedge products of the differentials of some
given (super)forms on the $(D + 2)$-dimensional supermanifold.
The usefulness of these parameters, in our whole discussion,  are primarily
four folds. First, these allow us, for instance, to take into account the fact
that there are three (super) differentials  corresponding to the 1-forms
defined on the four $(2 + 2)$-dimensional supermanifold. These are,
for the sake of emphasis, once again written as
$$
\begin{array}{lcl}
(dx^\mu), \qquad \;
(dx^\mu)\; s^{\theta\theta}, \qquad \;
(dx^\mu)\; s^{\bar\theta\bar\theta},
\end{array} \eqno(4.1)
$$
whose Hodge duals correspond to wedge products of the differentials
corresponding to the 3-forms on the supermanifold as given below
$$
\begin{array}{lcl}
&& \star\; (dx^\mu) = \varepsilon^{\mu\nu} (dx_\nu \wedge d \theta \wedge
d\bar\theta), \qquad
\star\; [(dx^\mu s^{\theta\theta})] = \varepsilon^{\mu\nu} (dx_\nu
\wedge d \theta \wedge d \theta), \nonumber\\
&& \star\; [(dx^\mu s^{\bar\theta\bar\theta})] = \varepsilon^{\mu\nu} (dx_\nu
\wedge d \bar \theta \wedge d\bar \theta).
\end{array} \eqno(4.2)
$$
In fact, the above prescription can be generalized to any $(D + 2)$-dimensional
supermanifold. Second, the presence of these parameters facilitate the action
of the double $\star$ operations on any arbitrary form $f$ that is supposed
to obey $\star (\star f) = \pm f$ [39]. For instance, in the above example,
the following results turn out automatically
$$
\begin{array}{lcl}
\star\; [\; \star\; (dx^\mu)\;] = d x^\mu, \quad
\star\; [\; \star\; (dx^\mu)\;s^{\theta\theta}\;]
= d x^\mu\; s^{\theta\theta}, \quad
\star\; [\; \star\; (dx^\mu)\;s^{\bar\theta\bar\theta}\;]
= d x^\mu\; s^{\bar\theta\bar\theta}.
\end{array} \eqno(4.3)
$$
Third, it is evident that the Hodge dual of a 2-superform (e.g. $d\theta \wedge
d \theta$) will be a 2-superform on a $(2 + 2)$-dimensional supermanifold.
The existence of $s^{\theta\theta}$ does allow such a definition because
$\star (d\theta \wedge d\theta) = \frac{1}{2!} \varepsilon^{\mu\nu}
(dx_\mu \wedge dx_\nu) s^{\theta\theta}$.
Fourth, the existence of the above parameters is at the heart of the
accurate derivation of the nilpotent (anti-)co-BRST symmetry transformations
for the gauge field as well as the (anti-)ghost fields
 of the 2D- and 4D free Abelian gauge theories
as is evident from the key equations (2.35), (2.36), (3.28) and (3.29).

It is clear from our present discussion that the geometrical
superfield formalism provides an exact and unique way of deriving
the local, covariant, continuous, nilpotent and anticommuting
(anti-)BRST symmetry transformations. However, this is {\it not}
the case with the derivation of the (anti-)co-BRST symmetries
which are not found to be {\it unique}. In fact, for both the 2D-
and 4D free 1-form Abelian gauge theories, the dual-horizontality
condition ($\tilde \delta \tilde A^{(1)} = \delta A^{(1)}$) leads
to the conditions $(\partial \cdot R) = 0, (\partial \cdot \bar R)
= 0, (\partial \cdot S) = 0$ on the secondary fields of the
expansions in (2.7) as well as (3.5). For the 2D theory, there
exist local, covariant, continuous and nilpotent solutions for
$R_\mu, \bar R_\mu, S_\mu$ so that one obtains the (anti-)co-BRST
transformations of (2.4). However, for the 4D free Abelian gauge
theory, only non-local, non-covariant, continuous and nilpotent
solutions exist for $R_\mu, \bar R_\mu, S_\mu$. Furthermore, these
solutions for the latter case are not {\it unique}. In fact, there
has been a whole lot of discussion on the various possibilities of
the existence of the dual-BRST symmetry transformations for the
Abelian gauge theory in [41]. All these possibilities of
symmetries are captured by different choices of $R_\mu$ and $\bar
R_\mu$ (see, e.g., [42] for details). Thus, in some sense, the
superfield formalism with the super co-exterior derivative $\tilde
\delta$ does provide the  reasons behind the non-uniqueness of the
nilpotent (anti-)co-BRST symmetry transformations where the
dual-horizontality condition plays a very decisive role.

It would be an interesting endeavour to generalize our present work
to the case of the interacting gauge theories where the gauge fields couple
to the matter fields. In fact, one such example,
where the $U(1)$ gauge field $A_\mu$ couples with the Dirac
fields in 2D, has been shown to present
the field theoretical model for the Hodge theory. In this model, the nilpotent
(anti-)BRST and (anti-)co-BRST symmetries co-exist together [33,34]. In
a recent set of papers [43-47],
the nilpotent symmetries for all the basic fields
of (i) the interacting
2D- as well as 4D (non-)Abelian gauge theories, and (ii) a
reparametrization invariant theory, have been derived by exploiting the
{\it augmented} superfield formulation. In this formalism, in addition
to the (dual-)horizontality conditions, the invariance of the (super)matter
conserved currents on the supermanifold has also been exploited. In fact, the
latter restriction yields the nilpotent symmetries for the matter fields
of an interacting gauge theory.
Furthermore, it would be an interesting
venture to generalize our present work to the
discussion of the free 4D 2-form Abelian gauge theory where the existence of
the local, covariant, continuous and nilpotent (anti-)co-BRST symmetries
has been shown [48,49]. Yet another direction that could be pursued
is to generalize the superfield formalism with only two Grassmann variables
(i.e. $\theta$ and $\bar\theta$) to the superfield approach
depending upon multiple Grassmann variables (e.g. $\theta_\alpha$ and
$\bar\theta_{\dot \alpha}$ with $\alpha, \dot \alpha = 1, 2, 3...$).
These are some of the issues that are under
investigation and our results would be reported elsewhere [50].\\

\noindent
{\bf Acknowledgements}\\

\noindent
Fruitful discussions with V. P. Nair, K. S. Narain and G. Thompson
at AS-ICTP, Trieste, Italy are gratefully acknowledged where
this work was initiated. Thanks are also due to the members of the
HEP group at AS-ICTP for their warm hospitality and useful conversations.

\end{document}